\documentclass{article}
\usepackage{fullpage}
\usepackage{tcolorbox}

\usepackage[comma]{natbib}

\usepackage{amsmath,amssymb,mathrsfs}
\usepackage{bm}
\usepackage{scalerel}
\usepackage{tikz}
\usetikzlibrary{arrows.meta}
\usetikzlibrary{hobby}
\definecolor{green}{rgb}{0.0,0.50,0.0}
\definecolor{blue}{rgb}{0.,0.,0.8}
\definecolor{almond}{rgb}{0.94,0.87,0.8}
\definecolor{brass}{rgb}{0.36,0.54,0.66}
\definecolor{olive}{rgb}{0.4,0.4,0}
\tikzset{>={Straight Barb[angle'=60, scale=1.1]}}

\def\bu{\mathbf{u}}
\def\bw{\mathbf{w}}
\def\bf{\mathbf{f}}
\def\bnua{\bm{\nu}_\mathrm{a}}
\def\nua{{\nu}_\mathrm{a}}
\def\bk{\mathbf{k}}
\def\bx{\mathbf{x}}
\def\bq{\mathbf{q}}

\def\barbu{\bar{\bar{\bu}}}
\def\baru{\bar{\bar{u}}}
\def\barphi{\bar{\bar{\varphi}}}

\def\barb{\bar{\bar{b}}}

\def\bnu{\bm{\nu}}
\def\blambda{\bm{\lambda}}
\def\balpha{\bm{\alpha}}
\def\p{\mathsf{p}}
\def\bpar{\bm{\partial}}

\def\bp{\bm{\p}}
\def\bz{\mathbf{z}}

\def\lie{\mathcal{L}}
\def\ip{\mathop{\lrcorner}}
\def\divv{\mathop{\mathrm{div}}}
\def\bv{\mathbf{v}}
\def\eps{\varepsilon}
\def\d{\mathrm{d}}
\def\C{\mathscr{C}}
\def\D{\mathscr{D}}

\def\p{\mathsf{p}}
\def\A{\mathsf{A}}
\def\B{\bm{\mathsf{B}}}

\newcommand{\barS}[1]{\overline{#1}^{{\mathrm{S}}}}
\newcommand{\barL}[1]{\overline{#1}^\mathrm{L}}
\newcommand\dt[2]{\frac{\d #1}{\d #2}}
\newcommand\ord[1]{^{(#1)}}
\newcommand\ordi[2]{^{(#1)#2}}

\newcommand{\half}{\tfrac{1}{2}}

\title{\textbf{Geometric Approaches to Lagrangian Averaging}}

\author{Andrew D. Gilbert$^1$ and Jacques Vanneste$^2$}
\date{\normalsize{$^1$Department of Mathematics and Statistics, University of Exeter, Exeter EX4 4QF, UK; email: A.D.Gilbert@exeter.ac.uk} \smallskip \\
\normalsize{$^2$School of Mathematics and Maxwell Institute for Mathematical Sciences, University of Edinburgh, Edinburgh EH9 3FD, UK; email: J.Vanneste@ed.ac.uk}}

\begin{document}



\maketitle

\noindent
Lagrangian averaging theories, most notably the Generalised Lagrangian Mean (GLM)  theory of \citet{andr-mcin78a},  have been primarily developed in Euclidean space and Cartesian coordinates. We re-interpret these theories using a geometric, coordinate-free formulation. This gives central roles to the flow map, its decomposition into mean and perturbation maps, and the momentum 1-form dual to the velocity vector. In this interpretation, the Lagrangian mean of any tensorial quantity is obtained by averaging its pull back to the mean configuration. Crucially, the mean velocity is not a Lagrangian mean in this sense. It can be defined in a variety of ways, leading to alternative Lagrangian mean formulations that include GLM and \citeauthor{sowa-robe10}'s  (\citeyear{sowa-robe10}) glm. These formulations share key features which the geometric approach uncovers. We derive governing equations both for the mean flow and for wave activities constraining the dynamics of the pertubations. The presentation focusses on the Boussinesq model for inviscid rotating stratified flows and reviews the necessary tools of differential geometry.  
\medskip



\section{Introduction}

Numerous fluid dynamical phenomena are the result of interactions between small-scale or high-frequency fluctuations -- associated with waves or turbulence -- and mean flows. Examples include acoustic streaming, the formation of jets in geophysical and astrophysical fluids, and the dynamo effect in conducting fluids.

Averaging, over time, space or ensembles of flow realisations, is central to the analysis of these phenomena since it enables the decomposition between mean flow and fluctuations that is required  both conceptually and practically. It has long been recognised that the most straightforward form of averaging, Eulerian averaging, that is averaging at fixed spatial location, often leads to unsatisfactory decompositions. This  especially true for advection-dominated, high-Reynolds-number flows whose  dynamics is controlled by material transport (of scalars such as temperature and potential vorticity, or of vectors such as  vorticity and magnetic field). Eulerian averaging does not preserve the structure of the advective terms in the equations governing fluid motion: the average of the material derivative along the flow differs from the material derivative along an averaged flow by terms -- Reynolds stresses or similar -- whose impact is often difficult to ascertain. As a result, the strong constraints that material conservation laws impose on the dynamics are obscured by the averaging process. 

Lagrangian averaging offers a solution. It replaces averaging at fixed position by averaging at fixed particle label or, in other words, averaging along fluid particle trajectories. Lagrangian averaging has a long history, dating back to pioneering work by \citet{ecka63,dewa70,bret71,sowa72,grim75} and others. The landmark paper by \citet{andr-mcin78a} provides solid foundations for Lagrangian averaging in the form of the theory of the Generalised Lagrangian Mean (GLM). 

A key idea of GLM and its predecessors is to use a suitably defined mean position of particles as a proxy for their label. This avoids the difficulties caused by the intricate nature of the mapping between particle labels and positions and gives the averaged equations of motion a familiar, Eulerian-looking appearance. \citet{andr-mcin78a} derive these averaged equations for a general compressible rotating fluid. 

There is now a large body of literature that applies these or similar GLM equations obtained for other fluid models to study the effect of waves on mean flows in the atmosphere and ocean \citep[e.g.][]{buhl-mcin98,buhl-mcin05,holm-et-al11,xie-v15}. A widespread application is to ocean surface waves. Their impact on currents is modelled using the Craik--Leibovich equations \citep{crai-leib76}, a version of the GLM equations specialised for high-frequency potential waves \citep{leib80,holm96,suzu-foxk}. The book by \citet{buhl14} gives a detailed account of GLM theory and its applications.

The derivation of GLM equations from the parent fluid model requires elaborate manipulations. The resulting equations have attractive properties but also disconcerting features. For instance, the key Lagrangian mean momentum equation (Theorem I of \citet{andr-mcin78a}, here ignoring the Coriolis terms) appears in the form
\begin{equation}
    (\partial_t + \barL{\bu} \cdot \nabla) (\barL{u}_i - \p_i) + \partial_i \barL{u}_k \, (\barL{u}_k - \p_k)  = - \partial_i ( \cdots),
    \label{eq:GLMThm1}
\end{equation}
where $\barL{\bu}$, in components $\barL{u}_i$, is the GLM mean velocity, $\p_i$ the components of the pseudomentum, and we do not detail the complicated pressure-like term on the right-hand side. The appearance of both $\barL{\bu}$ and $\barL{\bu} - \bp$ as velocity-like vectors is notable, as is the unusual contraction in the second term on the left-hand side.

In this review, we make the case that a geometric approach to Lagrangian averaging, using basic tools of exterior calculus, offers a straightforward route for the derivation and interpretation of GLM equations. 
The geometric viewpoint arises naturally in variational approaches to GLM \citep{grim84,gjaj-holm,holm02b,holm02a,salm98,salm13,salm16} but it is also useful when dealing directly with the equations of motion, as we do here.  At its most basic, a geometric approach makes it clear that Eq.\ \eqref{eq:GLMThm1} should be interpreted as 
\begin{equation}
    (\partial_t + \lie_{\barL{\bu}}) \barL{\bnu} = - d ( \cdots),
    \label{eq:GLMThm1nu}
\end{equation}
where $\barL{\bnu} = (\barL{\bu} - \bp) \cdot d \bx$ is a 1-form, $\lie_{\barL{\bu}}$ denotes the Lie derivative along $\barL{\bu}$ and $d$ is the differential. 
The GLM version of Kelvin's circulation theorem is then a short step away.
Beyond this, the geometric approach provides a unified way to understand a broad class of Lagrangian mean theories that generalize \citeauthor{andr-mcin78a}'s GLM. These theories differ in their definition of the Lagrangian mean flow but share many of GLM's properties, including the form of Eq.\ 
\eqref{eq:GLMThm1nu}. Lagrangian mean theories alternative to GLM have been proposed by \citet{robe-sowa06b, robe-sowa06a} and \citet{sowa-robe10,sowa-robe14} building on the pioneering work of \citet{sowa72}. The key feature of their so-called `glm' theory is that, for an incompressible fluid, the Lagrangian mean velocity is divergence free, unlike that of standard GLM \citep[see also][]{v-youn22}. \citet{gilb-v18} point out the underlying geometric foundations of Lagrangian mean theories and put forward alternatives to both GLM and glm.

A virtue of the geometric approach is that it leads to general results valid on arbitrary Riemannian manifolds and with arbitrary coordinate systems. This is useful for applications that include two-dimensional flows on the sphere or three-dimensional flows described in spherical coordinates. The geometric approach also sheds light on features of standard GLM that depend on the Euclidean structure assumed by \citet{andr-mcin78a} and \citet{buhl14}. The most prominent of such features is the definition of the Lagrangian mean velocity itself, which in GLM relies on the averaging of velocity vectors based at different points in space. Here, we follow \citet{gilb-v18} in giving the lead role to the mean flow map \cite[see also][]{holm02b,holm02a}. 
The mean flow map can be defined in a variety of ways, none of which should be thought of as resulting from the application of an averaging operator to flow maps (averaging operators involve summations and hence only apply to elements of vector spaces, which flow maps are not). The mean velocity is then simply the time derivative of the mean flow map. This viewpoint 
accommodates alternative Lagrangian mean theories (GLM, glm, etc.) in a unified way. 

This review focusses on two aspects of Lagrangian averaging: (i) the geometric derivation of the mean momentum equation, generalising Eq.\ \eqref{eq:GLMThm1nu}; and (ii) the derivation of wave activity conservation laws and the related equations governing the evolution of the pseudomomentum. Aspect (ii) gives a new geometric view of results obtained in the GLM framework by  \citet{andr-mcin78b} and \citet[][\S10.3]{buhl14} \citep[see also][]{grim84,salm13}. These results constrain the dynamics of the perturbations -- the difference between exact and mean fields that results for fluctuations such as waves -- and, via Eq.\ \eqref{eq:GLMThm1nu}, of the mean flow. 
We do not discuss the detailed models of  perturbations that are required to close Eq.\ \eqref{eq:GLMThm1nu}. These include  linear models for wave-like perturbations \citep[e.g.,][]{grim75,buhl-mcin98,gjaj-holm,buhl-mcin05,holm-et-al11,buhl14,wagn-youn,xie-v15,salm16, holm-hu-stree} and heuristic closures, either deterministic \citep[e.g.][]{holm99,mars-shko01,mars-shko03,robe-sowa09,sowa-robe08} or stochastic \citep[e.g.][]{holm19,holm21}. A geometric approach is also beneficial for these models, as demonstrated by the studies by Holm and co-workers on variational formulations.

We concentrate the presentation on the Boussinesq model of rotating stratified fluid \citep[e.g.][]{vall17}. This model is widely used in geophysical fluid dynamics. From the perspective of this review, it  has the advantage of illustrating the main features of a geometric approach to Lagrangian averaging. Extensions to fully compressible fluids or to magnetohydrodynamics are straightforward, if unwieldy \citep{gilb-v18,gilb-v21}.  We also restrict our attention to non-dissipative, unforced dynamics. Finally, we consider averaging as a general, abstract procedure, in the manner of \citet{andr-mcin78a} and most authors since. This is best thought of as ensemble averaging, that is, averaging over a collection of independent flow realisations. This can be approximated in practice by temporal or spatial averaging in a single flow, usually building on a time scale or spatial scale separation. 

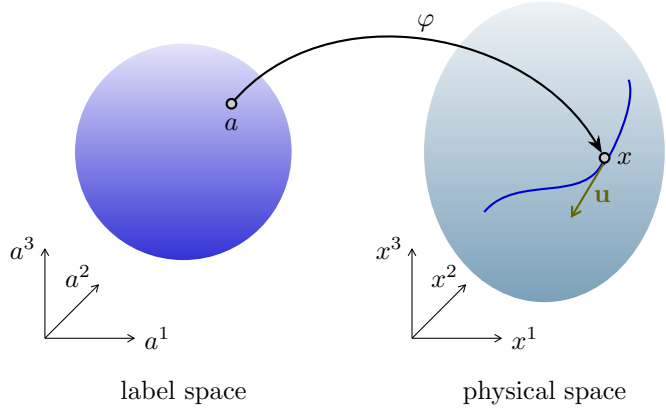
\begin{figure}
\centering
   \begin{tikzpicture}[scale=0.8,
    thick,
    dot/.style = {
      draw,
      fill = white,
      circle,
      inner sep = 0pt,
      minimum size = 4pt
    }
  ]
  \coordinate (O) at (0,0);
  \coordinate (A) at (0.8,0.8);
  \coordinate (B) at (7,-0.1); 
  \coordinate (C) at (6,-1);
  \coordinate (D) at (-2.3,-3.1);
  \coordinate (E) at (3.8,-3.1);

  \draw[->,thin] (D) -- ++(1.5, 0) node[right]{$a^1$};
  \draw[->,thin] (D) -- ++(0, 1.5) node[left]{$a^3$};
  \draw[->,thin] (D) -- ++(0.9, 0.9) node[left, xshift=0.05em,yshift=0.3em]{$a^2$};
  \shade[left color=blue!10, right color=blue!50, bottom color=blue!80] (O) ellipse (1.8 and 1.8);
  \draw[->,thin] (E) -- ++(1.5, 0) node[right]{$x^1$};
  \draw[->,thin] (E) -- ++(0, 1.5) node[left]{$x^3$};
  \draw[->,thin] (E) -- ++(0.9, 0.9) node[left, xshift=0.05em,yshift=0.3em]{$x^2$};
  \shade[left color=brass!10, right color=brass!50, bottom color=brass!80] (6,0) ellipse (2 and 2.5);
  \draw[-,blue] (B) to [out=250,in=50] (5.,-1);
  \draw[-,blue] (7.4,1.2) to [out=290,in=250] (B);
  \draw[->,olive] (7.04,-0.1) -- ++(-.6,-1) node[midway,xshift=.5em,yshift=-.3em]{$\bu$};

  \node at (0,-4) {label space};
  \node at (6,-4) {physical space};
  \draw[shorten >= 2pt,-{Stealth[scale=1]},thick,black] (A) to[out=50,in=120] node[above] {$\varphi$} (B); 
  \node[anchor=north,yshift=-0.2em,black] at (A) {$a$};
  \node[anchor=west,xshift=0.1em] at (B) {$x$}; 
  \draw[thick, black, fill=black!20] (A) circle (.8mm);
  \draw[thick, black, fill=black!20] (B) circle (.8mm);
 \end{tikzpicture}

    \caption{The flow map $\varphi$ maps the label $a=(a^1,a^2,a^3)$ of a fluid particle to the position $x=(x^1,x^2,x^3)$ of the particle at time $t$. The velocity field evaluated at $x$, $\bu(x)$, gives the velocity of the particle.}
    \label{fig:flowmap}
\end{figure}

A few comments about the notation used in this review are worthwhile. We have adopted as far as possible standard fluid-dynamical and GLM notation, such as boldface vectors and $\barL{\phantom{u}}$ to denote Lagrangian mean quantities, with a few exceptions. We use lightface fonts for the (Eulerian) position $x$ and label $a$ of fluid particles, with $x^i$ and $a^i$, $i=1,2,3$ as coordinates. This emphasise the distinction between points on a manifold and (tangent) vectors which can be overlooked in the Euclidean setting. The bold $\bx$ appears only as shorthand for $(x^1,x^2,x^3)$ and should be thought of as a collection of coordinates rather than a vector.
For convenience, we do not indicate the explicit time dependence of various objects. Thus, with $\varphi$ the (time-dependent) flow map, we write
\begin{equation}
    x = \varphi(a)
    \label{eq:flowmap}
\end{equation}
and
\begin{equation}
    \partial_t \varphi(a) = \bu(\varphi(a))
\end{equation}
instead of $x = \varphi(a,t)$ and $\partial_t \varphi(a,t) = \bu(\varphi(a,t),t)$. (See Figure \ref{fig:flowmap} for an illustration of the flow map and velocity vector.)
As is standard, we use the overbar $\overline{\phantom{\bu}}$ to denote (ensemble) average carried out at fixed independent variables, so $\bar{\bu} $ denotes the Eulerian mean velocity. 


We introduce an unconventional double bar notation for the mean flow map $\barphi$ and associated Lagrangian mean velocity $\barbu$. This emphasises that they are not the result of a direct averaging procedure. We restrict the use of the notation $\barL{\phantom{u}}$ for a specific definition of Lagrangian averaging (Eq.\ \eqref{eq:barL}) that applies to any tensorial object and on any manifold. It turns out that, even in standard GLM, the mean velocity is not consistent with this definition, so $\barL{\bu} \not= \barbu$. To follow the convention of the rest of this review, Eqs.\ \eqref{eq:GLMThm1} and \eqref{eq:GLMThm1nu} should therefore be rewritten with $\barbu$ in place of $\barL{\bu}$. We do not indicate explicitly the dependence of various fields on flow realisations. Instead, overbars of various kinds decorate all realisation-independent (mean) fields.

The differential geometry machinery we rely on is standard and exposed, for instance, in the books by \citet{schu80} or \citet{fran04}. The broad set up is a manifold -- the fluid domain -- equipped with a metric $g$, such  that $g(\bu,\bv) \in \mathbb{R}$ is the scalar or dot product of $\bu$ and $\bv$, and the associated volume form $\mu$. We use differential forms, mainly the momentum 1-form $\bnu$, vorticity 2-form $d \bnu$ and volume 3-form $\mu$, and operators including contraction $\ip$, exterior derivative $d$, Lie derivative $\lie$ and  push-forward and pull-back by maps. We summarise the relevant concepts in three sidebars. 
We primarily use a coordinate-free notation which greatly simplifies most derivations. The geometric approach is also effective when it comes to obtain explicit coordinate expressions. We illustrate this with a few detailed computations. 

\begin{tcolorbox}
\section*{Vectors and forms}
The familiar (contravariant) vectors  have duals, namely 1-forms (or covariant vectors) defined as linear maps on vectors, such that the pairing $ \balpha(\bv)$
of the 1-form $\balpha$ with the vector $\bv$ is a real number. In coordinates $x^i$, vectors and 1-forms are written as
\begin{equation}
\bv = v^i \bpar_i \quad \textrm{and} \quad \balpha = \alpha_i \, d x^i,
\end{equation}
where $\bpar_i = \mathbf{e}_i$ are basis vectors (the notation calls to mind the interpretation of vectors as directional derivatives) and $d x^i$ are  basis 1-forms, respectively, and summation is implied.  The two bases are dual in the sense that $d x^i (\bpar_j) = \delta^i_j$. Together with linearity, this implies the coordinate expression
\begin{equation}
    \balpha(\bv) = \alpha_i v^i
\end{equation} 
for the pairing between $\balpha$ and $\bv$. 

In three dimensions, there are three other differential forms: 0-forms, which are just scalars, and  2- and 3-forms. The 2-forms $\bm{\beta}$ and 3-forms $\bm{\gamma}$  are defined as bi- and tri-linear antisymmetric maps on vectors, with
\begin{equation}
\bm{\beta} = \tfrac{1}{2} \beta_{ij} \, d x^i \wedge d x^j \quad 
\textrm{and}  \quad \bm{\gamma} =  \tfrac{1}{6} \gamma_{ijk} \, d x^i \wedge d x^j \wedge d x^k.
\end{equation}
Here $\beta_{ij}$ and $\gamma_{ijk}$ are anti-symmetric in their indices and $(d x^i \wedge d x^j)(\bu,\bv) = u^{[i} v^{j]} = u^i v^j - u^j v^i$ and $(d x^i \wedge d x^j  \wedge d x^k)(\bu,\bv,\bw) = u^{[i} v^j w^{k]}$, where $[\cdots]$ denotes full (unnormalised) antisymmetrisation. 1-, 2- and 3-forms integrate over lines, surfaces and volumes, respectively. 

The interior product $\ip$ pairs a $k$-form with a vector to yield a $(k-1)$-form, with
\begin{equation}
    \bu \ip \balpha = \balpha(\bu) = \alpha_i u^i, \quad 
    \bu \ip \bm{\beta} =  u^i \beta_{ij} \, dx^j \quad \textrm{and} \quad
    \bu \ip \bm{\gamma} = \tfrac{1}{2} u^i \gamma_{ijk} \, dx^j \wedge d x^k. 
\end{equation}
The exterior derivative $d$ is a differential operator from fields of $k$-forms to fields of $(k+1)$-forms, with
\begin{equation}
  d f = \partial_i f \, d x^i, \quad   d \balpha = \partial_j \alpha_i \, d x^j \wedge d x^i, \quad
    d \bm{\beta} = \partial_k \beta_{ij} \, d x^k \wedge d x^i \wedge d x^j
\end{equation}
and $d \bm{\gamma}=0$. It satisfies $d^2 \equiv d \circ d =0$ and encodes gradient, divergence and curl. 
\end{tcolorbox}

\begin{tcolorbox}
\section*{Action of maps: push-forward and pull-back}
A map $\varphi$ such as the flow map (using the position at $t=0$ as label) takes a point $x$ in the fluid domain and sends it to another, $\varphi(x)$. Maps have a natural action on vectors and differential forms. A  vector $\bv$ at $x$ is \emph{pushed forward} to the vector $\varphi_* \bv$ at $\varphi(x)$ obtained by joining the images under $\varphi$ of the two infinitesimally close endpoints $x$ and $x + \eps \bv $  of $\bv$ with $\eps \to 0$. The \emph{push-forward} has coordinates
\begin{equation}
    (\varphi_* \bv)^i (\varphi(x))= v^j (x) \,\partial_j \varphi^i(x).
\end{equation}
Conversely, a 1-form $\balpha$ at $\varphi(x)$ is \emph{pulled back} to the 1-form $\varphi^* \balpha$ at $x$ in such a way that $\balpha(\varphi_* \bv) = (\varphi^* \balpha) ( \bv)$ for all $\bv$. In coordinates,
\begin{equation}
    (\varphi^* \balpha)_i (x) = \alpha_j (\varphi(x))\,\partial_i \varphi^j(x).
\end{equation}
For an invertible map, the pull-back of vectors and push-forward of 1-forms are defined as the push-forward and pull-back by the inverse map $\varphi^{-1}$. Push-forward and pull-back can be applied to any tensors, starting with scalar functions for which 
$(\varphi_* f)(x) = f(\varphi^{-1}(x))$
and 
$(\varphi^* f)(x) = f(\varphi(x))$ 
 The definition for higher-order tensors uses their interpretation as multilinear maps on vectors or 1-forms. A useful property is that push-forward and pull-back commute with the exterior derivative $d$.
\end{tcolorbox}

\begin{tcolorbox}
\section*{Lie derivative}
The notion of material derivative, or derivative along a flow, is expressed by the Lie derivative, an infinitesimal version of the pull-back which applies to scalars, vectors, 1-forms and higher-order tensors. Given a vector field $\bv$, we  can define the associated flow map $\psi_s$, such that 
\begin{equation}
    \dt{}{s} \psi_s(x) = \bv(\psi_s(x)) \quad \textrm{and} \quad \psi_0(x)=x,
    \label{eq:liederivdef}
\end{equation}
with $s$ a time-like variable (fictitious time), independent of the actual time $t$. The Lie derivative of a tensor field $\tau$ (independent of $s$) is then
\begin{equation}
    \lie_{\bv} \tau = \left. \dt{}{s} \right|_{s=0} \psi_s^* \tau.
        \label{eq:liederivdef1}
\end{equation}
This is the rate of change of $\tau$ as we pull it back under the flow  $\psi_s$ associated with $\bv$.
Applying this to scalar, vector and 1-form fields gives the coordinate expressions
\begin{equation}
    \lie_{\bv}\, f = v^j \partial_j f, \quad (\lie_{\bv}\, \bu)^i = v^j \partial_j u^i - u^j \partial_j v^i \quad \textrm{and} \quad (\lie_{\bv}\, \balpha)_i = v^j \partial_j \alpha_i + \alpha_j \partial_i v^j.
\end{equation}
The useful formula 
\begin{equation}
    \dt{}{t}\, \varphi^* \tau = \varphi^* \left(\partial_t + \lie_{\bu}\right) \tau
\end{equation}
relates the time derivative of the pull-back of any time dependent tensor $\tau$ by the flow map $\varphi$ to its Lie derivative with respect to the velocity field $\bu$. Kelvin's circulation theorem and the frozen-in nature of vorticity for the Euler equations are immediate applications of this formula.
The Lie derivative commutes with the exterior derivative operator $d$ and transforms naturally under pull-back:
\begin{align}
d \,  \lie_{\bv} \tau = \lie_{\bv} \, d \tau  \quad \textrm{and} \quad 
\varphi^* \lie_{\bv}\, \tau  = \lie_{\varphi^* \bv} \, \varphi^* \tau.
\end{align}
\end{tcolorbox}

\section{Geometric formulation of the Boussinesq equations}

We use the Boussinesq equations governing the dynamics of incompressible rotating stratified fluids as representative of the class of models typically considered in studies of wave--mean-flow interactions. In standard form, they read
\begin{subequations}
\label{eq:boussinesqu}
\begin{align}
\partial_t \bu + \bu \cdot \nabla \bu +  \bf \times \bu  &= - \nabla p + b \, \mathbf{e}_3, \label{eq:boussinesqu1} \\
\partial_t b + \bu \cdot \nabla b &= 0, \\
\nabla \cdot \bu & = 0,
\end{align}
\end{subequations}
where $\bu=(u^1,u^2,u^3)$ is the velocity vector, $\mathbf{e}_3$ is the vertical unit vector, $\bf = f \,\mathbf{e}_3$, with $f$ the Coriolis parameter, $p$ is the pressure scaled by the uniform background density $\rho_0$, and $b$ is the buoyancy acceleration, defined as $b = - g \rho / \rho_0$, with $\rho$ the density perturbation \citep[e.g.][]{vall17}. 

Eqs.\ \eqref{eq:boussinesqu} apply to Euclidean space and Cartesian coordinates. They generalise to arbitrary manifolds and coordinates in a way that is straightforward  in appearance only: the derivative $\bu \cdot \nabla \bu$ needs to be interpreted as the covariant derivative $\nabla_{\bu} \bu$, a sophisticated object that involves Christoffel symbols and is difficult to average. It is therefore preferable to write the Boussinesq equations in an alternative form that involves the  simpler Lie derivative -- the natural derivative along the flow. This necessitates making use of the momentum 1-form $\bnu = \nu_i \, d x^i$ as well as of the velocity vector $\bu = u^i \, \bpar_i$.
Here and in what follows, summation over repeated indices is understood.

For Eqs.\ \eqref{eq:boussinesqu}, $\bnu$ is just a version of $\bu$, obtained by lowering the superscript using the metric $g$,
\begin{equation}
    \bnu = \bu_\flat = g(\bu, \cdot), \quad \textrm{or, in coordinates,} \ \ \nu_i = g_{ij} u^j,
    \label{eq:nuu}
\end{equation}
In Cartesian coordinates, since $g_{ij} = \delta_{ij}$, this is simply
\begin{equation}
    \bnu = u_1 \, d x_1 + u_2 \, d x_2 + u_3 \, d x_3 =  \bu \cdot d \bx,
    \label{eq:nuucar}
\end{equation}
where  we use $\bx = (x^1,x^2,x^3)$ and the  dot product formally, as shorthand for handling multiple components, with no implication that $\bx$ or $d \bx$ should be regarded as vectors.

In terms of $\bnu$, Eqs.\ \eqref{eq:boussinesqu} take the form
\begin{subequations}
\label{eq:boussinesqnu}
\begin{align}
    (\partial_t + \lie_{\bu} ) (\bnu + \blambda) &= - d \pi + b \, d x^3, \label{eq:boussinesqnu1} \\
    (\partial_t + \lie_{\bu}) b & = 0, \label{eq:boussinesqnu2} \\
    \lie_{\bu}\,  \mu &= 0, \label{eq:boussinesqnu3}
\end{align}
\end{subequations}%
where 
$\lie_{\bu}$ denotes the Lie derivative along $\bu$ (see sidebar),
$\pi = p -  \tfrac{1}{2} | \bu |^2 - \blambda(\bu)$ with $| \bu |^2 = g(\bu,\bu) = \bnu(\bu)$, and 
\begin{equation}
\blambda = \tfrac{1}{2} (\bf \times \bx) \cdot d \bx= \half f (-x^2 \, dx^1 + x^1 \, dx^2).
\label{eq:lambda}
\end{equation}
The Coriolis terms can be verified to reduce to that in Eq.\ \eqref{eq:boussinesqu1} by making use of Cartan's formula to write 
\begin{equation}
\lie_{\bu} \blambda - d\blambda(\bu) = \bu \ip d \blambda = (\bf \times \bu) \cdot d \bx.
\end{equation}
The incompressibility condition \eqref{eq:boussinesqnu3} can equivalently be written as $\divv \bu = 0$ since $\lie_{\bu}\,  \mu = (\divv \bu ) \mu$ defines the divergence. 
 
The form of Eq.\ \eqref{eq:boussinesqnu}, sometimes referred to as Euler--Poincar\'e form, emerges from a variational formulation of the Boussinesq equations \citep{arno66,arno-kesh,holm-et-al98}. It applies to arbitrary manifolds possessing a metric $g$, with a suitable definition of $\blambda$ and the replacement of $d x^3$ in Eq.\ \eqref{eq:boussinesqnu1} by the differential of the geopotential.

The Lie derivative $\lie_{\bu}$ which appears in the momentum Eq.\ \eqref{eq:boussinesqnu1} is the natural derivative of the 1-form $\bnu$  along the flow of $\bu$. Unlike the covariant derivative of $\bu$ that appears in Eq.\ \eqref{eq:boussinesqu1}, it is independent of the metric and transforms straightforwardly under mappings. This makes Eqs.\ \eqref{eq:boussinesqnu} a convenient starting point for Lagrangian averaging.

For our purposes, it is essential to keep a clear distinction between the momentum 1-form field $\bnu$ and the vector field $\bu$, that is, between the advected and advecting fields in Eq.\ \eqref{eq:boussinesqnu1} For the Boussinesq equations and indeed for many other fluid models, they are directly related via Eq.\ \eqref{eq:nuu} This is not always the case, however. A simple example of a different, non-trivial relation between $\bnu$  and $\bu$ is provided by the hydrostatic Boussinesq equations: the hydrostatic approximation, which neglects vertical acceleration, amounts to replacing Eq.\ \eqref{eq:nuucar} by
\begin{equation}
    \bnu = u^1 \, d x^1 + u^2 \, d x^2, 
\end{equation}
while retaining the three components of $\bu$ elsewhere. Holm's $\alpha$-model \citep{holm99,mars-shko01,mars-shko03,oliv-vasy} or Tao's modification of the Euler equations \citep{tao16} are other, more involved examples of non-trivial relations between $\bnu$ and $\bu$. In the presence of rotation, it is convenient 
to rewrite Eq.\ \eqref{eq:boussinesqnu1} compactly as
\begin{equation}
    (\partial_t + \lie_{\bu} ) \bnua = - d \pi + b \, d x^3. \label{eq:boussinesqnua} 
\end{equation}
This introduces the absolute momentum
\begin{equation}
    \bnua = \bnu + \blambda, 
    \label{eq:nua}
\end{equation}
which is also not wholly related to $\bu$ via the metric.

\begin{tcolorbox}
\section*{Momentum equation: from 1-form to components}
We illustrate the direct manipulation of forms by deriving the Boussinesq momentum equation in coordinates from the 1-form formulation Eq.\ \eqref{eq:boussinesqnu1} The most useful rules for such manipulations are the Leibniz product rule for the Lie derivative and the commutation of Lie derivative and exterior derivative. With $\blambda = \lambda_i \, d x^i = \tfrac{1}{2} ( \bf \times \bx)_i \, d x^i$, we write the left-hand side of Eq.\ \eqref{eq:boussinesqnu1} as
\begin{subequations}
\begin{align}
    (\partial_t + \lie_{\bu}) (\nu_i \, d x^i + \lambda_i \, d x^i) &= \bigl((\partial_t + \lie_{\bu}) \nu_i \bigr) \, d x^i + \nu_i d (\lie_{\bu}\, x^i) +\lie_{\bu}(\lambda_i) \, d x^i+ \lambda_i \, d (\lie_{\bu} \,x^i) \\
    &=\bigl((\partial_t +u^j \partial_j) \nu_i \bigr) \, d x^i + \nu_i \, d u^i + u^j \partial_j \lambda_i \, d x^i+ \lambda_i  \, d u^i.
    \label{eq:1formtocomp1}
\end{align}
\end{subequations}
The right-hand side reads
\begin{equation}
- d ( p - \tfrac{1}{2} \nu_i u^i - \lambda_i u^i) + b \, d x^3 = -\partial_i p \, d x^i + \tfrac{1}{2} \nu_i \, d u^i + \tfrac{1}{2} u_i \, d \nu^i + \lambda_i \, d u^i + u^i \partial_j \lambda_i \, d x^j + b \, d x^3.
\label{eq:1formtocomp2}
\end{equation}  
Equating Eqs.\ \eqref{eq:1formtocomp1} and \eqref{eq:1formtocomp2} and noting that  $ u^j \partial_j \lambda_i \, d x^i - u^i \partial_j \lambda_i \, d x^j = (\bf \times \bu) \cdot d \bx$
and that, in Euclidean space, $\nu_i = u^i$  gives
\begin{equation}
 \bigl((\partial_t +u^j \partial_j) u_i + (\bf \times \bu)_i \bigr) \, d x^i  = - \partial_i p \, d x^i + b \, d x^3,   
\end{equation}
matching the standard form Eq.\ \eqref{eq:boussinesqu1}
of the Boussinesq momentum equation.
\end{tcolorbox}

\section{Lagrangian averaging}

Fluid models involve material transport, of momentum, vorticity and buoyancy in the case of the Boussinesq equations, as is made plain by the presence of the operator $\partial_t + \lie_{\bu}$ in Eqs.\ \eqref{eq:boussinesqnu} 
Lagrangian averaging preserves this structure by averaging various fields not at fixed Eulerian position $x$, but at fixed Lagrangian label $a$. We now review how this averaging is carried out.

\subsection{Scalar example: buoyancy equation}

Let us consider the (scalar) buoyancy $b$ as a straightforward example. Averaging Eq.\ \eqref{eq:boussinesqnu2} at fixed $x$, that is,  performing an Eulerian average, gives 
\begin{equation}
    \partial_t \bar{b} + \overline{\lie_{\bu}\,  b} =  \partial_t \bar{b} + \overline{u_j \partial_j b} = \partial_t \bar{b} + \lie_{\bar{\bu}}\,  \bar{b} + \overline{u'_j \partial_j b'} = 0, 
    \label{eq:eulerianb}
\end{equation}
where we have introduced perturbations to the mean fields, denoted by primes, by writing $\bu = \bar{\bu} + \bu'$ and $b = \bar{b} + b'$. The last term in Eq.\ \eqref{eq:eulerianb}, involving a product of perturbations, spoils the transport structure of the buoyancy equation. In contrast, if we let $B(a) = b(\varphi(a))$, Eq.\ \eqref{eq:boussinesqnu2} becomes $\partial_t B = 0$, leading upon averaging to
\begin{equation}
    \partial_t \bar{B} = 0,
\end{equation}
with $\bar{B}(a) = \overline{b(\varphi(a))}$ the Lagrangian average of the buoyancy.

While this equation is formally simple it is not practical: the intricate nature of the flow map in all but the simplest flows makes it impossible to recover useful information about $b$ from $B$ or $\bar{B}$. Generalised Lagrangian Mean (GLM) theories offer a solution by regarding $\bar{B}$ not as a function of the label $a$ but as a function of a mean position ascribed to the particle identified by $a$.

GLM relies on the decomposition
\begin{equation}
     \varphi = \Xi \circ \barphi
  \label{eq:mapdecom}
\end{equation}
of the flow map, factorised as the composition of the mean flow map $\barphi$ and a (realisation-dependent) perturbation map $\Xi$. The precise definition of $\barphi$ is unimportant for now. What matters is that it is chosen such that, in all flow realisations, the perturbations $\Xi$ remain close to the identity for long times. This is required so that $\barphi$ can serve as a representative of the ensemble of flow maps, in accordance with the intuitive interpretation of the mean. It is also necessary in practice to ensure the convergence of asymptotic or numerical procedures for the computation of Lagrangian mean quantities.

\begin{figure}
    \centering
    \begin{tikzpicture}[scale=0.8,
    thick,
    dot/.style = {
      draw,
      fill = white,
      circle,
      inner sep = 0pt,
      minimum size = 4pt
    }
  ]
  \coordinate (O) at (-2.5,0);
  \coordinate (A) at (-1.7,0.8);
  \coordinate (B) at (7,-0.1); 
  \coordinate (C) at (6,-1);
  \coordinate (D) at (-2.3,-3.1);
  \coordinate (E) at (3.8,-3.1);
  \coordinate (Bp) at (6.5,0);
  \coordinate (Bpp) at (7.2,.3);
  \coordinate (X) at (2.5,-2.5);
  
  \shade[left color=blue!10, right color=blue!50, bottom color=blue!80] (O) ellipse (1.8 and 1.8);
 
  \shade[left color=brass!10, right color=brass!50, bottom color=brass!80] (6,0) ellipse (2 and 2.5);
 
  \draw[->,olive] (B) -- ++(.6,-1) node[left,xshift=.5em,yshift=-.4em]{$\bu$};
  \draw[->,olive] (Bp) -- ++(.4,-1.2) ;
  \draw[->,olive] (Bpp) -- ++(.7,-.8) ;

  \shade[left color=brass!10, right color=brass!50, bottom color=brass!80] (1.75,-2) ellipse (2 and 2.5);

  \node at (-2.5,-2.5) {label space};
  \node at (6,3) {fluid domain};
  \node at (1.75,1) {mean domain};
  \draw[shorten >= 2pt,-{Stealth[scale=1]},thick,black] (A) to[out=50,in=120]  (B); 
  \draw[shorten >= 2pt,-{Stealth[scale=1]},thick,black] (A) to[out=40,in=130]  (Bp); 
  \draw[shorten >= 2pt,-{Stealth[scale=1]},thick,black] (A) to[out=60,in=120] node[above] {$\varphi$} (Bpp); 
  \draw[shorten >= 2pt,-{Stealth[scale=1]},thick,black] (A) to[out=350,in=120] node[below] {$\barphi$} (X); 
  \draw[shorten >= 2pt,-{Stealth[scale=1]},thick,purple] (X) to[out=350,in=260]  (B);
  \draw[shorten >= 2pt,-{Stealth[scale=1]},thick,purple] (X) to[out=360,in=260]  (Bp);
  \draw[shorten >= 2pt,-{Stealth[scale=1]},thick,purple] (X) to[out=340,in=290] node[below,xshift=-2em,yshift=-.7em] {$\Xi$} (Bpp);

  \node[anchor=east,yshift=-0.2em,black] at (A) {$a$};

  \draw[thick, black, fill=black!20] (A) circle (.8mm);
  \draw[thick, black, fill=black!20] (B) circle (.8mm);
  \draw[thick, black, fill=black!20] (Bp) circle (.8mm);
  \draw[thick, black, fill=black!20] (Bpp) circle (.8mm);
  \draw[thick, black, fill=black!20] (X) circle (.8mm);
  \draw[->,olive] (X) -- ++(.7,-.8) node[midway,xshift=-.7em,yshift=-.2em]{$\barbu$};
  
 \end{tikzpicture} 
    \caption{Decomposition of an  ensemble of flow maps $\varphi$, here represented by three realisations, into the mean map $\barphi$ and perturbation maps $\Xi$. The velocities $\bu$ of the fluid particle labelled by $a$ in each flow realisation, and its mean velocity $\barbu$, are also indicated.}
    \label{fig:decompostion}
\end{figure}
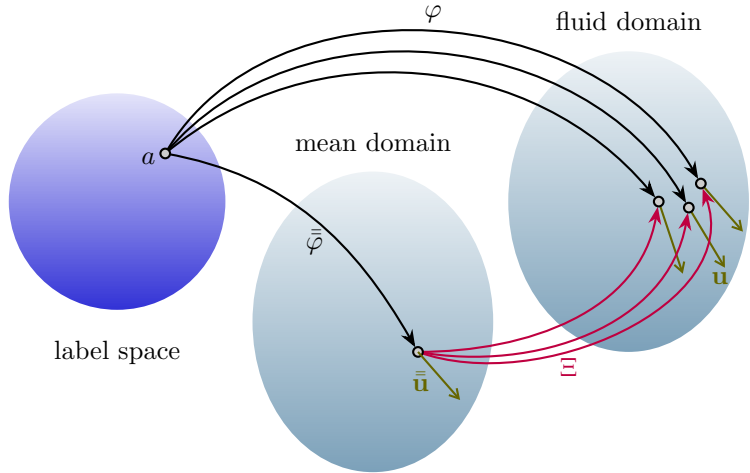

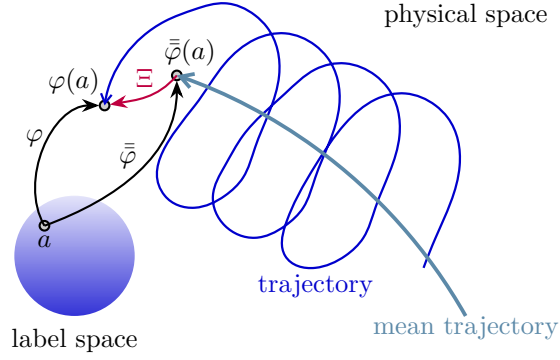
\begin{figure}
    \centering
   \begin{tikzpicture}[scale=0.8,
    thick,
    dot/.style = {
      draw,
      fill = white,
      circle,
      inner sep = 0pt,
      minimum size = 4pt
    }
  ]

\coordinate (A) at (-0.5,4.5);
\coordinate (B) at (0.7,5);
\coordinate (C) at (-1.5,2.5);

\shade[left color=blue!10, right color=blue!50, bottom color=blue!80] (-1,2) ellipse (1. and 1.);

\draw[thick, black, fill=black!20] (A) circle (.8mm);
  \draw[thick, black, fill=black!20] (B) circle (.8mm);
  \node[anchor=east,xshift=.2em,yshift=.9em,black] at (A) {$\varphi(a)$};
  \node[anchor=south,xshift=.5em,black] at (B) {$\bar{\bar{\varphi}}(a)$};
  \draw[thick, black, fill=black!20] (C) circle (.8mm);
  \node[anchor=north,xshift=.0em,black] at (C) {$a$};

 \draw [blue,<-,line width=0.3mm] plot [hobby] coordinates { (A) (.5,5.9) (1,6.1) (1.5,6.2) (2,6) (2.2,5.8) (2.4,5) (2,3.8) (1.8,3.3)
 (1.5,3) (1,2.8) (0.5,3)  (1,4.5) (1.5,5.2) (2,5.6) (2.5,5.7) (3,5.5) (3.2,5.3) (3.4,4.5) (3,3.3) (2.8,2.8)
 (2.5,2.5) (2,2.3) (1.5,2.5) (1.3,3.) (1.5,3.5) (1.8,4)
 (2.5,4.7) (3,5.1) (3.5,5.2) (4,5) (4.2,4.8) (4.4,4) (4,2.8) (3.8,2.3) 
 (3.5,2) (3,1.8) (2.5,2) (2.5,2.5)  (3,3.5) (3.5,4.2) (4,4.6) (4.5,4.7) (5,4.5) (5.2,4.3) (5.4,3.5) (5,2.3) (4.8,1.8)} ;
  
 \draw [brass,<-,line width=0.5mm] plot [hobby] coordinates { (B) (3,3.8) (5.5,1.)};
  
  \draw[shorten >= 2pt,-{Stealth[scale=1]},thick,black] (C) to[out=120,in=180] node[above,xshift=-.4em,yshift=-.3em] {$\varphi$} (A); 
  \draw[shorten >= 2pt,-{Stealth[scale=1]},thick,black] (C) to[out=20,in=270] node[above,xshift=-.4em,yshift=-.3em] {$\bar{\bar{\varphi}}$} (B);  
  
  \draw[shorten >= 2pt,-{Stealth[scale=1]},thick,purple] (B) to[out=230,in=10] node[above,xshift=0em,yshift=0em] {$\Xi$} (A);
  
   \node at (-1,0.6) {label space};
  \node at (5.5,6) {physical space};
  \node[brass] at (5.5,.8) {mean trajectory};
   \node[blue] at (3,1.5) {trajectory};

\end{tikzpicture}
    \caption{
    Alternative view of the flow map decomposition, in which the ensemble of flow maps arises from a phase shift in a nearly periodic motion. The mean map describes the `guiding-centre' trajectory of each fluid particle.}
    \label{fig:guidingcentre}
\end{figure}

Figure \eqref{fig:decompostion} visualises  the decomposition \eqref{eq:mapdecom} of the flow map, showing a small, three-member ensemble for illustration. An often-used ensemble arises for wave-like, nearly periodic motion: ensemble members are distinguished by a phase shift, and the mean map then traces the corresponding `guiding-centre' trajectory as illustrated in Figure \ref{fig:guidingcentre}

In general, $\barphi$ is not obtained by applying an averaging operator to the flow map, a notion that, in fact, makes little sense -- averaging is a linear operation, fundamentally an addition, but flow maps are not elements of a vector space and cannot therefore be added. We emphasise this with the (unconventional) double-bar notation. The Lagrangian mean velocity $\barbu = \baru^j \bpar_j$ is  defined as the  velocity field associated with $\barphi$, that is, such that
\begin{equation}
    \partial_t \barphi(a) = \barbu(\barphi(a)).   
\label{eq:baru}
\end{equation}
In general, the mean flow $\barbu$ is similarly not the result of the application of an averaging operator to the ensemble of vector fields $\bu$. Differentiating Eq.\ \eqref{eq:mapdecom} with respect to $t$ gives
\begin{equation}
    \bw + \Xi_* \barbu = \bu, \quad \textrm{i.e.} \quad  \barbu = \Xi^*(\bu - \bw),
    \label{eq:w}
\end{equation}
where $\bw$  is the perturbation velocity defined by $\partial_t \Xi(x) = \bw(\Xi(x))$ and we use that the inverse of the  push-forward $\Xi_*$ is the pull-back $\Xi^*$.
%

With this notation in place, the Lagrangian mean of $b$ is  defined by
\begin{equation}
    \barL{b}(\barphi(a)) = \bar{B}(a) = \overline{b(\varphi(a))}
\end{equation}
and satisfies
\begin{equation}
    \partial_t \barL{b} + \baru^j \partial_j \barL{b} = 0.
    \label{eq:barLbuoyancy}
\end{equation}
Eq.\ \eqref{eq:barLbuoyancy} clearly respects the transport structure of the original, unaveraged buoyancy equation, unlike Eq.\ \eqref{eq:eulerianb}

\subsection{General Lagrangian averaging and the momentum equation}

An alternative derivation of Eq. \eqref{eq:barLbuoyancy} uses the geometric machinery of pull-back and Lie derivative: first note that $\barL{b}(x)=\overline{b(\Xi(x))}$, that is,
\begin{equation}
    \barL{b} = \overline{\Xi^* b}.
    \label{eq:barLb}
\end{equation}
The key identity
\begin{equation}
    \Xi^* (\partial_t + \lie_{\bu}) \tau= (\partial_t + \lie_{\barbu}) \Xi^* \tau,
    \label{eq:keyidentity}
\end{equation}
valid for any tensor field $\tau$,  
can then be exploited \citep{gilb-v18}. Applying $\Xi^*$ to Eq.\ \eqref{eq:boussinesqnu2} gives, after averaging,
\begin{equation}
    (\partial_t + \lie_{\barbu}) \barL{b} = 0,
    \label{eq:LMbuoyancy}
\end{equation}
i.e.\ Eq.\ \eqref{eq:barLbuoyancy}

The advantage of this derivation is that it readily generalises: equations involving the material transport of an arbitrary tensor $\tau$ expressed as a Lie derivative are handled by first pulling them back to the mean configuration using $\Xi$, then averaging. This gives
\begin{equation}
   \overline{\Xi^* (\partial_t + \lie_{\bu}) \tau} = (\partial_t + \lie_{\barbu}) \barL{\tau}.
\end{equation}
Here, $\barL{\tau}$ denotes the Lagrangian mean of $\tau$,  defined as 
\begin{equation}
    \barL{\tau} = \overline{\Xi^* \tau},
    \label{eq:barL}
\end{equation}%
generalising the scalar definition of Eq.\ \eqref{eq:barLb} We emphasise that Eq.\ \eqref{eq:barL} provides a natural, coordinate-independent construction of the Lagrangian mean, one that applies to any tensorial object. It is geometrically sound, pulling back the ensemble of values of the tensor $\tau$ to the same point before carrying out the average. We reserve the notation 
$\barL{\phantom{u}}$ for this construction and caution that this convention is not adopted in standard GLM literature where $\barL{\phantom{u}}$ is sometimes used for mean fields constructed differently.

With the definition in Eq.\ \eqref{eq:barL}, applying $\Xi^*$ to Eq.\ \eqref{eq:boussinesqnua} yields
\begin{equation}
    (\partial_t + \lie_{\barbu}) 
    \Xi^*{\bnua} = - d \, \Xi^*{\pi} + \Xi^*({b \, d x^3}),
    \label{eq:Mmomentumpb}
\end{equation}
using that pull-back and exterior derivative $d$ commute.  Averaging then gives
\begin{equation}
    (\partial_t + \lie_{\barbu}) \barL{\bnua} = - d \barL{\pi} + \barL{b \, d x^3}.
    \label{eq:LMmomentum}
\end{equation}
The Lagrangian mean momentum equation \eqref{eq:LMmomentum} closely resembles the original momentum equation \eqref{eq:boussinesqnu1}. The key difference is that the link between  advected momentum and advecting velocity field -- the trivial $\bnu = \bu_\flat$ of the original equation -- is broken:  $\barL{\bnu}$ and $\barbu$ are not directly related, $\barL{\bnu} \not= \barbu_\flat$. Physically, this is because  perturbations to the mean have a dynamical effect which is manifested in the mismatch. It is  natural to introduce a field that quantifies the mismatch and hence encodes the impact of the perturbations on the mean. This field is the pseudomomentum (or at least a version thereof).

To include rotation, we define the pseudomomentum  1-form  $\bp$ as (minus) the difference between the absolute Lagrangian-averaged momentum $
\barL{\bnua} = \barL{\bnu} + \barL{\blambda}$ and the absolute momentum associated with the mean velocity, that is, 
\begin{equation}
    - \bp = \barL{\bnua} - (\barbu_\flat +  \blambda).
    \label{eq:pmdef1}
\end{equation}
With this definition, we can rewrite the Lagrangian-mean momentum Eq.\ \eqref{eq:LMmomentum} as
\begin{equation}
    (\partial_t + \lie_{\barbu})(\barbu_\flat - \bp) + \barbu \ip d \blambda= - d \bigl( \barL{\pi} + \blambda(\barbu) \bigr) +  \barL{b \, d x^3}.
    \label{eq:LMmomentum2}
\end{equation}
This is a geometric form of Theorem I of \citet{andr-mcin78a}. 

Several points are noteworthy. First, because the Lie derivative in Eq. \eqref{eq:LMmomentum2}  is applied to a 1-form, namely $\balpha = \barbu_\flat - \bp$, its coordinate expression reads $\baru^j \partial_j \alpha_i + \alpha_j \partial_i \baru^j$, which explains what might otherwise be thought as a peculiar combination of terms in Theorem I, as noted in Eq.\ \eqref{eq:GLMThm1}. Second, the Coriolis term 
$\barbu \ip d \blambda$
on the left-hand side involves the Lagrangian mean velocity rather than, say, the Lagrangian mean momentum or the Eulerian mean velocity. This is central to the conclusion that, for fast rotation, geostrophic balance is between the Lagrangian mean velocity and the gradient of a pressure-like mean field \citep{moor70,andr-mcin78a,xie-v15,wagn-youn,kafi-et-al21}. Third, while the pressure-like term is complicated,
\begin{equation}
\barL{\pi} + \blambda(\barbu) =  \overline{p \circ \Xi - \tfrac{1}{2}  (\bu \circ \Xi) \cdot \left(\bu \circ \Xi + \bf \times \bm{\Xi}  \right)} + \tfrac{1}{2} \barbu \cdot (\bf \times \bx),
\end{equation}
with $\bm{\Xi}=(\Xi^1,\Xi^2,\Xi^3)$,  its origin is clear. The same can be said of the coordinate expression for $\bp$, deduced from Eq.\ \eqref{eq:pmdef1} to be
\begin{equation}
    - \p_i = \overline{ (\nu_j \circ \Xi) \, \partial_i \Xi^j }
    - g_{ij} \baru^j + \tfrac{1}{2}  \overline{\left(  \bf \times \bm{\Xi} \right)_j \partial_i \Xi^j} - \tfrac{1}{2}(\bf \times \bx)_i .
    \label{eq:pseudomomentum}
\end{equation}

Fourth, the buoyancy term in Eq.\ \eqref{eq:LMmomentum2}, $\barL{b \, d x^3} 
= \overline{(\Xi^* b) \, d\, \Xi^3} 
= \overline{(b\circ\Xi) \,  d\, \Xi^3}$,
can be expressed in terms of the Lagrangian mean buoyancy under the assumption, which we make henceforth, that the ensemble of buoyancy fields to be averaged results from the same initial field, $b_0$ say. In this case, Eq.\ \eqref{eq:boussinesqnu2} implies that $b = \varphi_* b_0 = \Xi_* \barphi_* b_0$ hence 
\begin{equation}
    \Xi^* b = \barphi_* b_0
\end{equation} is independent of the flow realisation and equal to its average,
\begin{equation} 
 \Xi^* b  = \overline{\Xi^* b } = \barL b. 
 \label{eq:Xi*b}
\end{equation}
This reduces the buoyancy term in  Eq.\ \eqref{eq:LMmomentum2} to
\begin{equation}
   \barL{b \, d x^3} = \barL{b}  d \overline{\Xi^3}. 
   \label{eq:LMbuoyancysimple}
\end{equation}

\subsection{Circulation, vorticity and potential vorticity}

The circulation around a closed curve $\C$ is  expressed in terms of the momentum as
\begin{equation}
    \oint_\C \bnu,
\end{equation}
since 1-forms are naturally integrated along curves. The form of the momentum Eq.\   \eqref{eq:boussinesqnu1} is  ideal to derive Kelvin's circulation theorem. Let $\C_t = \varphi_* \C_0$ denote a material closed curve, thought of as the image at time $t$ of a  curve $\C_0$ in label space. Then,
\begin{subequations}
\begin{align}
    \dt{}{t} \oint_{\C_t} \bnua &= \dt{}{t}  \oint_{\C_0} \varphi^* \bnua = \oint_{\C_0} \varphi^* (\partial_t + \lie_{\bu}) \bnua 
    =  \oint_{\C_t} (\partial_t + \lie_{\bu}) \bnua \\
    &= \oint_{\C_t} b \, d x^3 = - \oint_{\C_t}  x^3 \, d b,
    \label{eq:circulation}
\end{align}
\end{subequations}
where we use Eq.\ \eqref{eq:boussinesqnua}, that $b \, d x^3 = d (b x^3) - x^3 \, d b$ and the fact that  integrals of exact differentials over closed curves vanish. The absolute circulation $\oint_{\C_t} \bnua$ is conserved provided that $\C_t$ lies on a constant-buoyancy surface $b = \mathrm{const}.$ (equivalently, $\C_0$ lies on a surface $b_0 = \mathrm{const}$).

The steps leading to Eq.\ \eqref{eq:circulation} can be applied to the Lagrangian mean momentum Eq.\ \eqref{eq:LMmomentum} For a contour $\overline{\C}_t = \barphi_* \C_0 = \Xi^* \C_t$ that moves with the mean velocity $\barbu$, this gives
\begin{equation}
    \dt{}{t} \oint_{\overline{\C}_t} \barL{\bnua} = \oint_{\overline{\C}_t} \barL{b \, d x_3} = - \oint_{\overline{\C}_t} \overline{\Xi^3} \, d  \barL{b},
\end{equation}
using Eq.\ \eqref{eq:LMbuoyancysimple} The same result is obtained by averaging Eq.\ \eqref{eq:circulation} directly, noting that
\begin{equation}
    \overline{\oint_{\C_t}\bnua}= \oint_{\overline{\C}_t} \overline{\Xi^* \bnua} =  \oint_{\overline{\C}_t} \barL{\bnua}.
\end{equation}

Vorticity is best considered as the 2-form $d \bnu$ obtained by taking the exterior derivative of the 1-form $\bnu$:
\begin{subequations}
\begin{align}
    d \bnu &= d ( \nu_1 \, d x^1 + \nu_2 \, d x^2 + \nu_3 \, d x^3) \\
    &= (\partial_2 \nu_3 - \partial_3 \nu_2) \, d x^2 \wedge d x^3  + (\partial_3 \nu_1 - \partial_1 \nu_3) \, d x^3 \wedge d x^1  + (\partial_1 \nu_2 - \partial_2 \nu_1) \, d x^1 \wedge d x^2.
    \label{eq:dnu}
\end{align}
\end{subequations}
In three dimensions, $d \bnu$ can be identified with a vorticity vector $\bm{\zeta}$ via $\bm{\zeta} \ip \mu = d \bnu$. (In $\mathbb{R}^3$, $d x^2 \wedge d x^3$ is identified with $\mathbf{e}_1 = \bpar_1$, etc., and the components in Eq.\ \eqref{eq:dnu} are also those of $\bm{\zeta}$.)
 The Boussinesq vorticity equation is a short step away from the momentum Eq.\ \eqref{eq:boussinesqnua}: applying $d$, using its commutation with the Lie derivative and that $d^2=0$ gives
\begin{equation}
    (\partial_t + \lie_{\bu}) d \bnua = d b \wedge d x^3.
    \label{eq:vorticity}
\end{equation}
In $\mathbb{R}^3$ and accounting for incompressibility, the corresponding vector form $(\partial_t + \lie_{\bu}) \bm{\zeta}_\mathrm{a} = \partial_2 b \, \mathbf{e}_1 - \partial_1 b \, \mathbf{e}_2$, involves the familiar transport term $\bu \cdot \nabla \bm{\zeta} - \bm{\zeta} \cdot \nabla \bu$.

The material conservation of potential vorticity is deduced by taking the wedge product of Eq.\ \eqref{eq:vorticity} with $d b$, which satisfies $(\partial_t + \lie_{\bu}) d b = 0$, and using that $d b \wedge d b =0$ to obtain the material conservation law
\begin{equation}
    (\partial_t + \lie_{\bu}) Q = 0,
    \label{eq:PV}
\end{equation}
where $Q$ is the potential vorticity 3-form
\begin{equation}
    Q = d \bnua \wedge d b.
\end{equation}
As a 3-form, $Q$ is naturally integrated over volumes. It can be interpreted as \citeauthor{hayn-mcin87}'s (\citeyear{hayn-mcin87,hayn-mcin90}) `potential vorticity substance', with its nature as a divergence made clear by rewriting it as $Q = - d ( b\, d \bnua )$ \citep[see also][]{bret-scha93}. The usual scalar potential vorticity $q$ is defined by 
\begin{equation}
    q \mu = Q.
\end{equation}
It is also conserved, $(\partial_t + \lie_{\bu}) q = 0 $, thanks to the incompressibility condition $\lie_{\bu} \, \mu = 0$.

The Lagrangian mean counterparts of Eqs.\ \eqref{eq:vorticity} and \eqref{eq:PV} are readily obtained by pulling back these equations and averaging. With $\Xi^* b = \overline{\Xi^* b} = \barL{b}$, we obtain from Eq.\ \eqref{eq:vorticity} that
\begin{equation}
    (\partial_t + \lie_{\barbu}) \barL{d \bnua} = d \barL{b} \wedge d \overline{\Xi^3}, \quad \textrm{where} \quad   \barL{d \bnua} = \overline{\Xi^* d \bnua} = d \barL{\bnua}
\end{equation}
is both the Lagrangian mean of the vorticity, and the vorticity associated (via the exterior derivative, equivalent to a curl) with the Lagrangian mean momentum. Similarly, Eq.\ 
 \eqref{eq:PV} gives
 \begin{equation}
     (\partial_t + \lie_{\barbu}) \barL{Q} = 0, \quad \textrm{where} \quad \barL{Q} = \overline{\Xi^* Q} = d \barL{\bnua} \wedge d \barL{b}
     \label{eq:LMPV}
 \end{equation}
is both the Lagrangian mean of the potential vorticity, and the potential vorticity constructed from  $\barL{\bnua}$ and $\barL{b}$.

In summary, provided that the Lagrangian mean momentum 1-form $\barL{\bnua}$ is taken as starting point instead of the mean velocity vector $\barbu$ (or the vector $\barbu + \tfrac{1}{2} \bf \times \bx$), the relations between  Lagrangian mean momentum, vorticity and potential vorticity are straightforward and parallel those of the original Boussinesq equations. The differential geometric machinery makes the derivation of these relations particularly simple, relying on the consistent use of Eq.\ \eqref{eq:barL} to define all Lagrangian mean quantities and on the commutation of the exterior derivative with the pull-back $\Xi^*$ and Lie derivative.  

A subtlety arises with the scalar potential vorticity. This is because the Lagrangian mean velocity $\barbu$ need not be 
divergence free  or, equivalently, the mean map $\barphi$ need not be
volume preserving, unlike the divergence free $\bu$ and volume preserving $\varphi$ in each flow realisation (we discuss this further in \S\ref{sec:means}). 
To handle this, we define the Lagrangian mean mass 3-form
\begin{equation}
    \barL \mu = \Xi^* \mu  = \barphi_* \mu.
\end{equation}
The second equality follows from the incompressibility condition $\varphi_* \mu = \mu$. It shows that  $\Xi^* \mu$ is realisation independent and hence equal to its average $\barL{\mu}$. In Euclidean space, with $\mu = d x^1 \wedge d x^2 \wedge d x^3$,
\begin{equation}
     \barL \mu = d\Xi^1 \wedge d\Xi^2 \wedge d \Xi^3 = \det \bigl(\partial_j \Xi^i \bigr) \, \mu.
\end{equation}
Pulling back the continuity equation with $\Xi^*$ shows that $\barL \mu$  satisfies
\begin{equation}
    (\partial_t + \lie_{\barbu}) \barL{\mu} = 0.
    \label{eq:liemuL}
\end{equation}
Combining this with Eq.\ \eqref{eq:LMPV} gives the material conservation along the mean flow $\barbu$ of the scalar $\bar{\bar{q}}$ defined by $\barL{\mu} \bar{\bar{q}} = Q$. This differs from the Lagrangian mean $\barL{q}$ of $q$ unless $\barphi$ preserves volume. The effective mean density $\tilde \rho$ used by \citet{andr-mcin78a}, \citet{buhl14} and others to handle the divergence of the mean flow is such that $\tilde \rho \mu = \barL \mu$.

\section{Mean flow definitions} \label{sec:means}

So far, we have not specified how the mean velocity $\barbu$ is chosen beyond the loose requirement that the associated flow map $\barphi$ be a good representative of the ensemble of flow maps $\varphi$. This makes it clear that the results of the previous sections are independent of this choice. The conventional definition of $\barbu$ and $\barphi$ is that of the GLM theory of \citet{andr-mcin78a}. The simplest way of expressing this definition is as follows. For a choice of coordinates $x^i$, the flow map has coordinates $\varphi^i(a)$. The mean flow map is defined coordinate-wise by 
\begin{equation}
    \barphi^i(a) = \overline{\varphi^i(a)}.
    \label{eq:GLM1}
\end{equation}
Thus the mean position of particle $a$ at time $t$ is obtained by averaging the three coordinates of that particle at time $t$ over flow realisations. Equivalently, we can write this as $\overline{\Xi^i(x)} = x^i$ or, introducing the displacement components $\xi^i$ such that  
\begin{equation}
    \Xi^i(x)=x^i + \xi^i(x),
     \label{eq:GLM2}
\end{equation}
as
\begin{equation}
     \overline{\xi^i(x)} = 0.
      \label{eq:GLM3}
\end{equation}
This implicit form is the one originally proposed by \citet{andr-mcin78a} and used since. We emphasise that the construction is based on a choice of coordinates. Only in Euclidean space, where positions can be interpreted as vectors, have Eqs. \eqref{eq:GLM1}--\eqref{eq:GLM3}  an intrinsic vectorial meaning.

The GLM mean velocity is obtained by differentiating Eq.\ \eqref{eq:GLM1} with respect to $t$ to find $\baru^i(\barphi(a)) = \overline{u^i(\varphi(a))}$, hence
\begin{equation}
    \baru^i(x) = \overline{u^i(\Xi(x))}, \quad \textrm{i.e.} \quad \baru^i = \overline{u^i \circ \Xi}.
    \label{eq:GLMbaru}
\end{equation}
In general, this expression is not geometrically intrinsic: it depends on the choice of coordinates in which the components $u^i$ are expressed. The tempting coordinate-free version $\barbu = \overline{\bu \circ \Xi}$ is not well defined: the vectors $\bu \circ \Xi$ corresponding to different flow realisations live in different tangent planes and hence cannot be averaged, unless in Euclidean space where vectors can be  translated freely. 

The lack of intrinsic geometric meaning results in inconvenient properties.
First, the Lagrangian mean velocity is generally divergent, that is, the Lagrangian mean flow does not preserve volume, unlike all the original flows in the ensemble \citep{mcin88}. Second, the range of $\barphi$ may differ from the fluid domain, most strikingly if the domain is a spherical shell or the surface of a 2-sphere. On the other hand, Eqs.\ \eqref{eq:GLM1} and \eqref{eq:GLM3} lead to simplifications. For instance, the pseudomomentum in Eq.\ \eqref{eq:pseudomomentum} reduces to
\begin{equation}
    - \p_i = \overline{ (u^j \circ \Xi ) \, \partial_i \xi^j } + \tfrac{1}{2} \overline{(\bf \times \bm{\xi})_j \, \partial_i \xi^j}, 
    \label{eq:GLMp}
\end{equation}
when we further assume the space to be Euclidean so that $g_{ij}=\delta_{ij}$. The explicit definition of the GLM mean flow is also convenient for the numerical computation of Lagrangian averages \citep{kafi22,kafi-v23}.

An alternative definition of the mean flow is the glm definition proposed by  \citet{sowa-robe10}. It replaces Eq.\ \eqref{eq:GLM3} by a geometrically meaningful condition. The central idea is to use a (realisation-dependent) vector field, $\bq$ say, as a proxy for the perturbation map $\Xi$ ($\bq$ should not be confused with the scalar potential vorticity $q$ introduced earlier).  The vector field $\bq$ generates $\Xi$ in the sense that the flow of $\bq$ from some fictitious time $\eps=0$ to $\eps=1$ is $\Xi$. Explicitly, $\bq$ defines a one-parameter family of flow maps $\Xi_\eps$ (at fixed $t$) with
\begin{equation}
    \partial_\eps \Xi_\eps(x) = \bq(\Xi_\eps(x)), \quad  \quad \Xi_0(x)=x \ \ \textrm{and} \ \ \Xi_1(x) = \Xi(x).
    \label{eq:Xiq}
\end{equation}
The glm mean flow is then prescribed by requiring that
\begin{equation}
    \overline{\bq} = 0.
\end{equation}
This is a geometric statement that applies on any manifold and is independent of the choice of coordinates. Importantly, for incompressible fluids, the volume preservation of $\Xi$ and hence of $\barphi$ is easily enforced through the linear constraint 
\begin{equation}
    \divv \bq = 0,
\end{equation}
which survives averaging. A disadvantage of glm is that the property $\overline{\xi^i}=0$ does not hold, which precludes simplifications such as those leading to the compact form Eq.\ \eqref{eq:GLM3} Another is that the correspondence between $\bq$ and $\Xi$ is only formal: not all diffeomorphisms can be obtained as time-$1$ flows of a steady vector field. It can however be implemented using asymptotic expansions, disregarding questions of convergence.  

Definitions of the mean flow other than those of GLM or glm are  possible. \citet{gilb-v18} propose to take $\barphi$ as the map closest to the realisations $\varphi$ in a least-square sense, with the distance between flow maps taken as the `Arnold distance' that is, the distance whose geodesics are solutions of the Euler equations \citep{arno66,ebin-mars,arno-kesh,holm-et-al}. Different definitions lead to a different decomposition of the flow maps $\varphi$ between mean part $\barphi$ and perturbations $\Xi$. All acceptable definitions should lead to mean flow maps 
that remain close to one another for long times.

\section{Small-amplitude perturbations}

Practical results are  obtained using perturbation expansions based on the closeness of the perturbation maps $\Xi$ to the identity. A typical  set up expands the velocity as 
\begin{equation}
    \bu = \eps \bu\ord{1} + \eps^2 \bu\ord{2} + \cdots,
\end{equation}
where $\bu\ord{1}$ is a wave field with $\overline{\bu\ord{1}}=0$ and $\eps \ll 1$ an amplitude parameter. Here and in what follows $\cdots$  denotes neglected terms that are $O(\eps^3)$.
In Euclidean space and with the GLM definition, the mean velocity is
\begin{align}
    \barbu &= \overline{\bu(x + \eps \xi\ord{1} + \eps^2 \xi\ord{2} +  \cdots)} \nonumber  \\
    &= \underbrace{\eps^2 \overline{{\bu\ord{2}}}}_{\textrm{Eulerian mean}} + \underbrace{\eps^2 \overline{\bm{\xi}\ord{1} \cdot \nabla \bu\ord{1}}}_{\textrm{Stokes drift } \barS{\bu}} + \cdots .
    \label{eq:smallGLM1}
\end{align}

The Stokes drift and hence Lagrangian mean velocity are divergent, with
\begin{equation}
    \divv \barbu = \partial_i \baru^i = \eps^2 \,\partial_t \bigl( \tfrac{1}{2} \partial_{ij}  \bigl(\overline{\xi\ordi{1}{i}\xi\ordi{1}{j}} \bigr) \bigr) + \cdots,
\end{equation}
where we use that $\partial_t \bm{\xi}\ord{1} = \bu\ord{1}$. Because this is an exact time derivative, the volume change associated with this divergence stays bounded and small over long times.

The Lagrangian mean momentum is
\begin{align}
    \barL{\nua}_i &= \overline{\left(\nu_j \circ \Xi + \tfrac{1}{2} (\bf \times \bm{\Xi})_j \right) \partial_i \Xi^j } \nonumber \\ 
    &=  \tfrac{1}{2}  (\bf \times \bx)_i + \eps^2 \overline{{\nu}_i\ord{2}}  
     + \eps^2 \, \overline{\xi\ordi{1}{j} \partial_j \nu\ord{1}_i + \nu\ord{1}_j \partial_i  \xi\ordi{1}{j} + \tfrac{1}{2} \left( \bf \times \bm{\xi}\ord{1} \right)_j  \partial_i \xi\ordi{1}{j} } + \cdots.
    \label{eq:smallGLM2}
\end{align}
In Euclidean space, $u^i = \nu_i$ and we can freely raise or lower indices. The pseudomomentum is then 
\begin{equation}
    - \p_i = \eps^2 \, \overline{\bigl(u\ordi{1}{j} \partial_i  \xi\ordi{1}{j} + \tfrac{1}{2} ( \bf \times \bm{\xi}\ord{1} )_j \partial_i \xi\ordi{1}{j}  \bigr)} + \cdots,
    \label{eq:smallGLM3}
\end{equation}
consistent with Eq.\ \eqref{eq:GLMp}
Eqs.\ \eqref{eq:smallGLM1}--\eqref{eq:smallGLM3} are standard results of GLM theory. Together with the Lagrangian mean momentum equation, they make it possible to predict the evolution of the Lagrangian and Eulerian mean flows when the wave field $\bu\ord{1}$ is prescribed. 

To obtain their glm counterparts, we implement a perturbative approach that uses $\Xi_\eps$ in Eq.\ \eqref{eq:Xiq} rather than $\Xi_1$ as the perturbation map. We use the formal representation
\begin{equation}
    \Xi_\eps^* = \exp \left( {\int^\eps_0 \lie_{\bq} \, \d \eps} \right) = \mathrm{id} + \eps \lie_{\bq\ord{1}} + \tfrac{1}{2} \eps^2 \left( \lie_{\bq\ord{1}}^2 + \lie_{\bq\ord{2}}  \right) + \cdots
    \label{eq:lieseries}
 \end{equation}
of the pull-back by $\Xi$ in terms of the generating vector field $\bq$ \citep[e.g.][]{lich-lieb,v-youn22}. In particular, 
\begin{equation}
    \Xi^i(x) = \Xi^* x^i = x^i + \eps q\ordi{1}i + \tfrac{1}{2} \eps^2 \bigl( q\ordi{1}j \partial_j q\ordi{1}i +  q\ordi{2}i \bigr) + \cdots,
\end{equation}
so that
\begin{equation}
    \bm{\xi}\ord{1} =  \bq\ord{1} \quad \textrm{and} \quad 
     {\bm{\xi}}\ord{2} =  \tfrac{1}{2} \bigl( \bq\ord{1} \cdot \nabla \bq\ord{1} +  \bq\ord{2} \bigr).
     \label{eq:xipert}
\end{equation}
Thus, at a linear, $O(\eps)$ level, the displacement $\bm{\xi}\ord{1}$ and the generating vector $\bq\ord{1}$ can be identified. The constraint $\overline{\bq} =0$  implies the non-zero quadratic mean displacement 
\begin{equation}
    \overline{\bm{\xi}\ord{2}} = \tfrac{1}{2} \overline{\bm{\xi}\ord{1} \cdot \nabla \bm{\xi}\ord{1}}.
    \label{eq:glmxi2}
\end{equation} 
This stands in contrast with the GLM defining assumption of zero mean displacements, Eq.\ \eqref{eq:GLM3}.

Time differentiating Eq.\ \eqref{eq:xipert} further gives the perturbation velocity  $\bw =( \partial_t \Xi)\circ \Xi^{-1}$ as 
\begin{equation}
    \bw = \eps \,\partial_t \bq\ord{1}+ \tfrac{1}{2} \eps^2 \bigl( \partial_t \bq\ord{2} - \lie_{\bq\ord{1}}  \partial_t \bq\ord{1} \bigr) + \cdots .
\end{equation}
Using Eq.\ \eqref{eq:lieseries} in Eq.\ \eqref{eq:w} leads to
\begin{equation}
    \barbu = \eps \bigl(\bu\ord{1} - \partial_t \bq\ord{1}\bigr) + \eps^2 \bigl( \bu\ord{2} - \tfrac{1}{2} \partial_t \bq\ord{2} + \lie_{\bq\ord{1}} \bu\ord{1} - \half \lie_{\bq\ord{1}} \partial_t \bq\ord{1}\bigr) + \cdots .
\end{equation}
The $O(\eps)$ term needs to vanish for the left-hand side to be a mean quantity, hence $\partial_t \bq\ord{1} = \bu\ord{1}$. Taking this into account in the $O(\eps^2)$ term and averaging leads to
\begin{equation}
    \barbu = \underbrace{\eps^2 \,\overline{\bu\ord{2}}}_{\textrm{Eulerian mean}} + \underbrace{\tfrac{1}{2} \eps^2 \,\overline{\lie_{\bq\ord{1}} \bu\ord{1}}}_{\textrm{solenoidal Stokes drift }\barS{\bu}_\textrm{sol}} + \,\cdots.  
\end{equation}
This is divergence free, in contrast to the GLM equivalent in Eq.\ \eqref{eq:smallGLM1}
\citet{v-youn22} argue in the context of surface waves that the glm, solenoidal Stokes drift, which in Eucidean space can be rewritten as
\begin{equation}
    \barS{\bu}_\mathrm{sol} = \tfrac{1}{2} \eps^2 \,  \overline{\bq\ord{1} \cdot \nabla \bu\ord{1} - \bu\ord{1} \cdot \nabla \bq\ord{1}} = \tfrac{1}{2} \eps^2 \, \nabla \times \bigl(\overline{\bu\ord{1} \times \bq\ord{1}} \bigr),
\end{equation}
is an advantageous alternative to the traditional Stokes drift $\barS{\bu} = \eps^2 \overline{\bm{\xi}\ord{1} \cdot \nabla \bu\ord{1}}$.
The difference,
\begin{equation}
 \barS{\bu} -  \barS{\bu}_\mathrm{sol} =   \eps^2\,\bigl(\overline{\bm{\xi}\ord{1} \cdot \nabla \bu\ord{1} - \tfrac{1}{2} \bm{\xi}\ord{1} \cdot \nabla \bu\ord{1} + \tfrac{1}{2} {\bu}\ord{1} \cdot \nabla \bm{\xi}\ord{1}} \bigr) = \tfrac{1}{2} \eps^2 \,\partial_t \bigl(\overline{\bm{\xi}\ord{1} \cdot \nabla \bm{\xi}\ord{1}}\bigr)
\end{equation}
(using that $\bq\ord{1}=\bm{\xi}\ord{1})$, is a time derivative and hence stays bounded over long time scales. This confirms that the mean trajectories according to GLM and glm definitions stay close to one another. The glm Lagrangian mean momentum is obtained directly using Eq.\ \eqref{eq:lieseries} as 
\begin{equation}
    \barL{\bnua} =  \tfrac{1}{2} ( \bf \times \bx) \cdot d \bx +\eps^2 \,\overline{\bnu\ord{2}} + \eps^2 \bigl( \overline{\lie_{\bq\ord{1}} \bnu\ord{1} + \tfrac{1}{4} \lie_{\bq\ord{1}}^2 ( \bf \times \bx) \cdot d \bx } \bigr) + \cdots.
\end{equation}

We note that, regardless of the specific definition of the mean, the Lagrangian mean of the velocity in the sense of Eq.\ \eqref{eq:barL} is  approximated as  
\begin{equation}
    \barL{\bu} = \overline{\Xi^* \bu} = \eps^2 \,\overline{\bu\ord{2}} + \eps^2 \,\overline{\lie_{\bq\ord{1}} \bu\ord{1}} + \cdots.
\end{equation}
This differs from the mean velocity fields $\barbu$ obtained for GLM and glm, by an $O(\eps^2)$ term that is not an exact time derivative. Thus the integral curves of $\barL{\bu}$ drift away from those of $\barbu$ (the mean trajectories $\barphi$), by $O(1)$ distances after long, $O(\eps^{-2})$ times leading to perturbation maps $\Xi$ well away from the identity. This makes $\barL{\bu}$ so defined unsuitable as a mean velocity. 

For simplicity, we have assumed in the above that the velocity $\bu$ has no $O(1)$ term. \citet{gilb-v18} give more general results that include the contributions from an $O(1)$, realisation-independent velocity $\bu\ord{0} = \overline{\bu\ord{0}}$. 

\section{Wave action} \label{sec:action}

 We now turn to the geometric formulation of wave activity conservation laws.  Wave activities are quantities that, in the small-amplitude limit, are quadratic in the perturbation fields and so can be approximated to leading order, that is, to $O(\eps^2)$, from knowledge of an $O(\eps)$ approximation to the perturbation fields. Their conservation constrains the dynamics of the perturbations and can serve as a basis to parameterise wave--mean flow interactions. We use geometric language to extend  results of \citet{andr-mcin78b} and \citet[][\S10.3]{buhl14} beyond the Euclidean setting, highlighting the role of symmetries.
 
 We begin with the conservation of wave action, the wave activity associated with phase averaging. Suppose that the fluctuation maps $\Xi$ in the decomposition Eq.\ \eqref{eq:mapdecom} depend periodically on a phase parameter $\alpha$. An ensemble of flows is then obtained by varying $\alpha$. The associated average corresponds to integration with respect to $\alpha$, so that any average of an $\alpha$ derivative vanishes, $\overline{\partial_\alpha \,\cdot\,} = 0$. 
 Naturally the metric $g$ and volume form $\mu$ do not depend on $\alpha$ and so for example we can write
\begin{align}
   \partial_\alpha  g(\bu,\bu) = \partial_\alpha (\bnu(\bu)) = 2 g(\bu, \partial_\alpha \bu) = 2 \bnu(\partial_\alpha \bu). 
\label{eq:unuuseful}  
\end{align}
The mean map $\barphi$ is also independent of $\alpha$. With this we can differentiate the composition of maps $\varphi = \Xi \circ \barphi$ with respect to time $t$ or $\alpha$ and obtain two vector fields: $\bu = (\partial_t \varphi) \circ \varphi^{-1}$, as earlier, and $\bz = (\partial_\alpha \varphi) \circ \varphi^{-1} = (\partial_\alpha \Xi) \circ \Xi^{-1}$. These vector fields may be related through equating mixed second derivatives to obtain the key link
\begin{align}
    \partial_\alpha \bu  = \partial_t \bz + \lie_{\bu} \, {\bz}.
    \label{eq:keylierelation}
\end{align}

Our starting point in deriving the wave action equation is the momentum equation \eqref{eq:boussinesqnu1}, which we first contract with $\bz$. Let us focus first on the key term $(\partial_t + \lie_{\bu})\bnua $ to obtain
\begin{subequations}
\begin{align}
   [ (\partial_t + \lie_{\bu} )\bnua](\bz) 
   &= (\partial_t + \lie_{\bu}) (\bnua(\bz)) - \bnua((\partial_t + \lie_{\bu})\bz ) 
   \label{eq:waveactivity1}  
   \\
  &= (\partial_t + \lie_{\bu}) (\bnua(\bz)) - \bnua(\partial_\alpha\bu ) 
     \label{eq:waveactivity2}\\
  &= (\partial_t + \lie_{\bu}) (\bnua(\bz)) -\partial_\alpha \left(  \half \bnu(\bu) + \blambda(\bu) \right) ,
     \label{eq:waveactivity3}
\end{align}
\end{subequations}
making use of Eqs.\ \eqref{eq:unuuseful} and \eqref{eq:keylierelation}
This is  written as transport of $\bnua(\bz)$ and an $\alpha$-derivative that  vanishes under averaging \citep{gilb-v18}.  
However the scalar $\bnua(\bz)$ does not provide a satisfactory definition of wave activity for the full equation of motion since its approximation to $O(\eps^2)$ requires evaluation of the vector field $\bz$ to $O(\eps^2)$.  A more suitable definition \citep[following][]{andr-mcin78b} is obtained by  applying the pull-back $\Xi^*$  and using Eq.\ \eqref{eq:keyidentity} to find
\begin{subequations}
\begin{align}
 \Xi^*\bigl([  (\partial_t + \lie_{\bu} )\bnua](\bz) \bigr)
   & =
\Xi^* \left[(\partial_t + \lie_{\bu}) (\bnua(\bz)) \right]
-  \Xi^* \partial_\alpha \left(  \half \bnu(\bu) + \blambda(\bu)\right) 
\\
& = (\partial_t + \lie_{\barbu}) \Xi^*(\bnua(\bz)) 
-   \partial_\alpha h
+(\partial_\alpha\Xi^* ) 
\left( \half \bnu(\bu) + \blambda(\bu)\right), 
\end{align}
\end{subequations}
where $h = \Xi^*(  \half \bnu(\bu) + \blambda(\bu))$ and is unimportant as $\partial_\alpha h$  vanishes under averaging. 
Now the derivative of the pull-back $\partial_\alpha \Xi^*$ is linked to the Lie derivative by $\bz$ according to 
\begin{align}
    (\partial_\alpha \Xi^*)\tau = \Xi^* \lie_{\bz} \, \tau
    \label{eq:alphaderivuseful}
\end{align}
for any tensor $\tau$, generalising Eq.\ \eqref{eq:liederivdef1} Thus we obtain 
\begin{align}
 \Xi^*\bigl([ (\partial_t + \lie_{\bu} )\bnua](\bz) \bigr)
   &=
\  (\partial_t + \lie_{\barbu}) \Xi^*(\bnua(\bz)) 
-   \partial_\alpha h
+ \Xi^* \lie_{\bz}\,  ( \half \bnu(\bu) + \blambda(\bu)).
  \label{eq:waveactivity10}
\end{align}

Returning to Eq.\ \eqref{eq:boussinesqnu1}, for the pressure-like term we have
$\Xi^*(- d\pi(\bz)) =  - \Xi^* \lie_{\bz} \pi $, which combines with the last terms of Eq.\ \eqref{eq:waveactivity10} to give $- \Xi^* \lie_{\bz} p $, with $p$ the original pressure.
For the buoyancy term $b\, dx^3$, contracting with $\bz$ and pulling back by $\Xi^*$ gives
\begin{align}
\Xi^* [ (b\, dx^3)(\bz)]  & = (\Xi^* b) \, \Xi^* [ dx^3(\bz)] = \barL b \, \Xi^* \lie_{\bz}\,  x^3
 = \barL b \,  \partial_\alpha ( \Xi^*  x^3) = \partial_\alpha ( \barL b \Xi^3),
   \label{eq:waveactivity11}
\end{align}
making use of Eq. \eqref{eq:Xi*b} and again of Eq.\ \eqref{eq:alphaderivuseful}
The definition of the wave activity as $\Xi^*(\bnua(\bz))$, instead of $\bnua(\bz)$, is key to Eq.\ \eqref{eq:waveactivity11}  which reduces the buoyancy term to an $\alpha$-derivative, to vanish on averaging.

Putting all this together, from applying $\Xi^*$ to the contraction of Eq.\ \eqref{eq:boussinesqnu1} with $\bz$, we  obtain%
\begin{align}
(\partial_t + \lie_{\barbu}) \Xi^*(\bnua(\bz)) 
-  \partial_\alpha \bigl( h  + \barb\, \Xi^3  \bigr)
 =  - \Xi^* \lie_{\bz} \, p .
 \label{eq:waveactivity15}
 \end{align}
We now average to eliminate the second term on the left-hand side, giving
\begin{align}
(\partial_t + \lie_{\barbu}) \barL{\bnua(\bz)}
  =  - \barL{\lie_{\bz} \, p }.
 \label{eq:waveactivity15a}
 \end{align}
We then multiply by $\barL \mu = \Xi^* \mu$, and simplify using Eq.\ \eqref{eq:liemuL}, Cartan's formula and 
$\barL{\lie_{\bz} \, p} 
\,\barL{\mu} 
=  d (\barL{p\bz} \ip \barL{\mu})
$, 
which follows from writing
\begin{equation}  
   \Xi^*(\lie_{\bz} \,p) \barL \mu = \Xi^*\bigl( (\lie_{\bz} \,p) \mu \bigr) = \Xi^* \bigl( \lie_{\bz} ( p \mu) \bigr)
    = \lie_{\Xi^* \bz} \bigl((\Xi^* p)  \barL \mu\bigr) = d \bigl(\Xi^*(p \bz) \ip \barL \mu\bigr)
\end{equation}
and averaging. 
This leads to a conservation law with local form
\begin{equation}
    \partial_t ( \A \, \barL \mu) + d \bigl((\A \barbu + \B) \ip \barL \mu \bigr) = 0,
    \label{eq:wavecons0}
\end{equation}
where we have defined the wave action density  $\A$ and non-advective flux $\B$ as 
\begin{equation}
    \A = \overline{\Xi^* ( \bnua(\bz)) } = \barL{\bnua(\bz)} \quad \textrm{and} \quad
    \B =\overline{ \Xi^* (p\bz) }  = \barL{p\bz}.
        \label{eq:wavecons1}
\end{equation}
The corresponding global form is deduced by integrating over a domain $\D$ with boundary $\partial \D$ to obtain
\begin{align}
   \frac{d}{dt} \int_{\D} \A \, \barL \mu + \int_{\partial \D} (\A \barbu + \B) \ip \barL \mu = 0 ,
            \label{eq:wavecons2}
\end{align}
using Stokes theorem. 
%

Eqs.\ \eqref{eq:wavecons0}--\eqref{eq:wavecons1} give a coordinate-free version of the wave action density and flux derived by \citet{andr-mcin78b} and \citet{buhl14}. We obtain coordinate expressions matching theirs by noting that 
\begin{subequations}
\begin{align}
\A  &= 
\overline{\Xi^* ( {\nua}_i \, z^i) } 
= \overline{({\nua}_i \circ \Xi) \,  (z^i \circ \Xi)}
= \overline{({\nua}_i \circ \Xi) \, \partial_\alpha\Xi^i} \label{eq:A} \\
\textrm{and} \quad
\mathsf{B}^i &= \overline{(p \circ \Xi) \, (\Xi^* \bz)^i} = \overline{(p \circ \Xi) \, (\partial_j \Xi^i)^{-1} \, (z^j \circ \Xi)} = \overline{(p \circ \Xi) \, (\partial_j \Xi^i)^{-1} \, \partial_\alpha \Xi^j},
\end{align}
\end{subequations}
where $(\partial_j \Xi^i)^{-1}$ denotes the $(i,j)$ entry of the inverse of the matrix $\partial_j \Xi^i$ and can be expressed in terms of the co-factors of this matrix.

We check that the wave action is a wave activity by showing that $\A$ and $\B$  can be calculated to leading order
solely from the leading order perturbation fields. 
With the expansions
\begin{equation}
\Xi^i(x) = x^i + \eps \xi\ordi{1}{i} + \eps^2 \xi\ordi{2}{i} + \cdots \quad \textrm{and} \quad
\bnua = \bnua\ord{0} + \eps \bnua\ord{1} + \eps^2 \bnua\ord{2}  + \cdots,
\end{equation}
where  $\bnua^{(0)}  = \overline{\bnua^{(0)}}$ is independent of $\alpha$, we compute
\begin{equation}
\partial_\alpha \Xi^i =\eps \partial_\alpha \xi\ordi{1}{i} + \eps^2 \partial_\alpha \xi\ordi{2}{i}  + \cdots \quad \textrm{and} \quad
\bnua \circ \Xi = \bnua\ord{0} + \eps \bnua\ord{1}  +  \eps \bm{\xi}\ord{1} \cdot \nabla  \bnua\ord{0}  + O(\eps^2), 
\end{equation}
Introducing this into Eq.\ \eqref{eq:A} and noting that the average removes 
$\eps \nu_{\mathrm{a}i}^{(0)}  \partial_\alpha \xi\ordi{1}{i}$ 
and, for GLM,
$\eps^2 \nu_{\mathrm{a}i}^{(0)} \partial_\alpha\xi^{(2)i} $, 
leaves
\begin{align}
\mathsf{A}  &=  \barL{\bnua(\bz)}  = \eps^2\,   \overline{\nu_{\mathrm{a}i}^{(1)}\,  \partial_\alpha\xi^{(1)i}} + \eps^2\, \partial_j \nu_{\mathrm{a}i}^{(0)}\,  \overline{  \xi\ordi{1}{j}  \, \partial_\alpha\xi\ordi{1}{i}}  + \cdots.
\end{align}
For glm, the term $\eps^2 \overline{\nu_{\mathrm{a}i}^{(0)} \partial_\alpha\xi^{(2)i}}$ persists, but Eq.\ \eqref{eq:glmxi2} can be used to express it in terms of the first order field $\bm{\xi}\ord{1}$. The $O(\eps^2)$ leading order approximation to $\B$ can be derived similarly.

\section{Symmetry and Killing vectors}

We have developed theory so far in a geometric setting, making use of a general metric $g$ and associated volume form $\mu$, although for simplicity we have used an Euclidean formulation of the Coriolis term and buoyancy force. It is often natural to single out a preferred direction, in particular in order to consider a mean flow $\barbu$ and ensemble of perturbations $\Xi$ that are invariant under translations in this direction. The Euclidean setting of the original GLM theory gives a privileged role to the directions along the three Cartesian coordinates. This is reflected in many of its results, especially those framed in terms of energy--momentum tensors \citep{andr-mcin78b,grim84,salm13}.
In spherical geometry, the longitude $\phi$ is typically singled out, leading to the consideration of axisymmetric  mean flows and  zonal averaging.

To develop theory along these lines, and capture the notion of some ignorable coordinate geometrically, we introduce the notion of a Killing vector field $\bk$ \citep{schu80,fran04}. This has the properties 
\begin{align}
    \lie_{\bk}\, g = 0 \quad \textrm{and} \quad \lie_{\bk}\, \mu = 0  
    \label{eq:killing}
\end{align}
(the second follows from the first). 
The first property can also be expressed as
\begin{align}
    \nabla \bk_\flat + (\nabla \bk_\flat )^\intercal = 0  \quad \text{or} \quad \nabla_i k_j + \nabla_j k_i = 0 , 
\end{align}
where $\nabla$ denotes the covariant derivative and $k_i = (\bk_\flat)_i$,
and so $\bk$ is familiar to fluid dynamicists as a vector field with zero rate-of-strain tensor: the flow of $\bk$ does not distort fluid elements. We denote by $\psi_\alpha$ the flow map obtained by integrating along $\bk$ for a fictitious time $\alpha$. In Cartesian geometry we could take, for example, $\bk = \bpar_1 $ and $\psi_\alpha$ is then the translation $x^1 \mapsto x^1 + \alpha$ through $\alpha$ along the $x^1$-axis. In spherical geometry, typically $\bk = \bpar_\phi$ and the transformation $\psi_\alpha$ is the rotation $\phi \mapsto \phi +\alpha$. 
With a Killing vector field $\bk$, we can express the invariance of the mean flow $\barbu$ in this direction by 
$\lie_{\bk}\, \barbu = 0$, 
and likewise for any other tensor.

Let us first make use of a Killing field to obtain a conserved momentum.  Contracting Eq.\ \eqref{eq:boussinesqnua} with $\bk$ yields, using Eq.\ \eqref{eq:killing} and standard manipulations,
\begin{equation}
       (\partial_t + \lie_{\bu} ) (\bnua(\bk) ) - (\lie_{\bk}\,\blambda)(\bu) = - \lie_{\bk}\,  p + b \, dx^3(\bk),
       \label{eq:consmomkilling} 
\end{equation}
where $dx^3(\bk)=k^3$, the vertical component of $\bk$.
To gain a conservation law from this we need to remove the Coriolis and buoyancy terms $\lie_{\bk}\,\blambda$ and $b k^3$. The latter vanishes provided that $\bk$ lies wholly in the $(x^1,x^2)$ directions.
If $\bk$ represents a rotation about the $x^3$ axis then $\blambda \propto \bk_\flat$ and the Coriolis term also vanishes. However if $\bk$ is a translation, that is  constant $k^1$ and $k^2$, then it does not, with $\lie_{\bk} \,\blambda = \half f(-k^2\, dx^1 + k^1\,dx^2)$. In this case we can use the gauge freedom discussed in \citet[][\S10.4.4]{buhl14}: since the Coriolis term in Eq.\ \eqref{eq:boussinesqnu1} depends on $d \blambda$ rather than $\blambda$ (as shown using Cartan's formula), $\blambda$ is defined up to the transformation $\blambda \mapsto \blambda + d h$ for any scalar function $h$. 
For a translation, say $\bk = \bpar_1$, we can pick $h = - \half f d(x^1 x^2)$ leading to the new, equivalent Coriolis term $\blambda  = - fx^2\, dx^1$
which satisfies $\lie_{\bk} \, \blambda = 0 $ as required. 

With the Coriolis and buoyancy terms removed by whatever means, 
integrating Eq.\ \eqref{eq:consmomkilling} times $\mu$ over a domain $\D$
gives
\begin{equation}
    \frac{d}{dt}
    \int_{\D} \bnua(\bk)\,  \mu 
       + \int_{\partial \D}  \bigl(\bnua(\bk)  \bu + p \bk\bigr) \ip \mu = 0 , \label{eq:consmom} 
\end{equation}
which is conservation, in each realisation in the ensemble, of linear momentum if $\bk$ generates translations and angular momentum if $\bk$ generates rotations. 

The evolution of momentum on Lagrangian parcels in Eq.\ \eqref{eq:consmomkilling} can also be pulled back to the mean flow by applying $\Xi^*$, to obtain 
\begin{equation}
       (\partial_t + \lie_{\barbu} ) \Xi^*(\bnua(\bk) ) - \Xi^* \! \left(\lie_{\bk}(\blambda)(\bu) \right) = - 
       \Xi^* \lie_{\bk}\,  p + \barL{b} \, \Xi^*(dx^3(\bk)),
       \label{eq:consmomkilling2} 
\end{equation}
using that $\Xi^*b = \barL{b}$.
Averaging then gives
\begin{equation}
    (\partial_t + \lie_{\barbu} ) \barL{\bnua(\bk)} - \barL{\lie_{\bk}(\blambda)(\bu)} = - \barL{\lie_{\bk}\, p} + \barL{b} \, \barL{d x^3(\bk)} .
    \label{eq:a1}
\end{equation}
When, as above, the Coriolis and buoyancy terms are removed, this corresponds to a conservation law, namely
\begin{equation}
        \frac{d}{dt}
        \int_{\D} \barL{\bnua(\bk)}\,  \barL{\mu}
       + \int_{\partial \D}  \bigl(\barL{\bnua(\bk)}  \barbu + \barL{p \bk}\bigr ) \ip \barL{\mu} = 0. 
       \label{eq:consmompullback} 
\end{equation}
obtained by multiplying by $\barL{\mu}$ and integrating over $\D$.

On the other hand, we can contract the Lagrangian mean momentum equation \eqref{eq:LMmomentum} with $\bk$ to obtain
\begin{equation}
    (\partial_t + \lie_{\barbu}) \barL{\bnua}(\bk) + \barL{\bnua} \!\left(\lie_{\bk} \,\barbu\right) = - \lie_{\bk} \,\barL \pi + \barL{b} d \barL{x^3}(\bk).
    \label{eq:a2}
\end{equation}
This leads to the conservation law 
\begin{equation}
    \frac{d}{dt}
        \int_{\D} \barL{\bnua}(\bk) \,  \barL{\mu}
       + \int_{\partial \D}  \barL{\bnua(\bk)} \barbu  \ip \barL{\mu} = 0 ,
       \label{eq:consmompullback1}
\end{equation}
provided that the mean fields are invariant to translation along the Killing vector field $\bk$ (since $d \barL{x^3}(\bk) = \lie_{\bk} \barL{x^3}$), as is the case for spatial averaging in the direction of $\bk$.

Subtracting Eq.\ \eqref{eq:a2} from Eq.\ \eqref{eq:a1} leads to the equation
\begin{equation}
    (\partial_t + \lie_{\barbu} ) \p_{\bk} - \barL{\lie_{\bk}(\blambda)(\bu)}   - \barL{\bnua} \!\left(\lie_{\bk} \,\barbu\right) = - \barL{\lie_{\bk}\, p} + \lie_{\bk} \,\barL \pi + \barL{b} \, \bigl(\barL{d x^3(\bk)} - d \barL{x^3}(\bk) \bigr)
    \label{eq:oo}
\end{equation}
for the wave activity
\begin{align}
 -\p_{\bk}  = \barL{\bnua}(\bk) - \barL{\bnua(\bk)}.
    \label{eq:pmdef2}
\end{align}
This wave activity can be interpreted as a form of pseudomomentum, related to but distinct from the general pseudomomentum defined in Eq.\ \eqref{eq:pseudomomentum} We emphasise the differences: $\p_{\bk}$ is a scalar defined for a specific Killing field $\bk$, whereas $\bp$ is a 1-form defined independently of such a field. Assuming that $\blambda(\bk)=0$, the natural relation
\begin{equation}
    \bp(\bk) = \p_{\bk}
\end{equation}
only holds provided that $\barL{\bnu(\bk)} = \barbu_\flat(\bk)$, which clearly depends on the definition chosen for the mean flow. 
In Euclidean space, we can take $\bk = \bpar_i$ for $i=1,2,3$. 
With the GLM definition of the mean, we then have
\begin{equation}
\barL{\bnu(\bpar_i)} = \barL{\nu_i} = \overline{\nu_i \circ \Xi} = \overline{u^i \circ \Xi } = \baru^i.
\end{equation}
As a result, the two versions of the pseudomomentum coincide:
\begin{equation}
    \p_i = \p_{\bk} \quad\text{for}\quad \bk = \bpar_i.
\end{equation}
This is an advantage of the GLM definition of the mean flow which is tied to the coordinate representation of the flow map and velocity field or, in the language of this section, to a choice of three independent Killing fields.

Eq.\ \eqref{eq:oo} reduces to the (Boussinesq form of the) GLM pseudomomentum equation \citep[][Eq.\ (10.125)]{buhl14}. When $\blambda(\bk)=0$, $k^3=0$ and the mean flow is invariant under translation along $\bk$ it yields the conservation law (the difference of Eqs.\ \eqref{eq:consmompullback} and \eqref{eq:consmompullback1}), 
\begin{equation}
    \frac{d}{dt}
        \int_{\D} \p_{\bk} \,  \barL{\mu}
       + \int_{\partial \D}  \bigl(\p_{\bk} \barbu + \barL{p \bk} \bigr)  \ip \barL{\mu} = 0.
\end{equation}

We conclude by observing the relation between the construction leading to Eq.\ \eqref{eq:pmdef2} and the derivation of the wave action in \S\ref{sec:action}. Given a Killing field $\bk$ and an ensemble of perturbation maps $\Xi$, we can generate a broader ensemble of maps $\Xi_\alpha$ 
by conjugation under the translation $\psi_\alpha$, 
\begin{equation}
   \Xi_\alpha = \psi_{-\alpha} \circ \Xi \circ \psi_{\alpha}.
\end{equation}
We can associate with this the $\alpha$-independent vector field 
\begin{equation}
    \bz =  \left. (\partial_\alpha \Xi_\alpha)\circ \Xi_\alpha^{-1} \right|_{\alpha=0}
    = \Xi_* \bk - \bk.
\end{equation}
The action in Eq.\ \eqref{eq:wavecons1} corresponding to this is 
\begin{equation}
\A = \overline{\Xi^* (\bnua(\bz))} = \overline{(\Xi^*\bnua)(\bk)} - \overline{\Xi^* (\bnua(\bk))} = \barL{\bnua}(\bk) - \barL{\bnua(\bk)} = - \p_{\bk} .
\end{equation}
Eq.\ \eqref{eq:oo} for $\p_{\bk}$ is closely related to wave action conservation, with some differences that disappear when the averaging is along the direction of $\bk$. In variational formulations of Lagrangian averaging \citep{grim84,holm02b,holm02a,salm13}, the conservation of $\p_{\bk}$ in this case is interpreted as arising from Noether's theorem.

\section{Summary points}
\begin{enumerate}
\item Lagrangian averaging is best formulated starting with a geometric formulation of fluid models in terms of the momentum 1-form $\bnu$. 
\item When decomposing flow maps as $\varphi = \Xi \circ \barphi$, various choices can be made for the mean map $\barphi$ and corresponding mean velocity $\barbu$. \citeauthor{andr-mcin78a}'s GLM and \citeauthor{sowa-robe10}'s glm theories correspond to two distinct choices.
\item The Lagrangian mean of any tensor $\tau$ is  defined as $\barL{\tau} = \overline{\Xi^* \tau}$. The Lagrangian mean momentum $\barL{\bnu}$ is the natural dynamical variable for the mean equations. Its difference with the 1-form $\barbu_\flat$ associated with $\barbu$, defined as pseudomomentum, encodes the impact of perturbations on the mean.
\item Geometrically intrinsic forms of wave activities are readily derived. The conservation law of wave action arises for an ensemble of perturbations parameterised by a phase; further conservation laws are generated by Killing vector fields leaving the mean fields invariant.
\item The approach is general and easily handles different coordinate systems, for example spherical polar coordinates and axisymmetric fields, and different manifolds, for example flow on the surface of a 2-sphere.
\end{enumerate}

\section*{Disclosure statement}
The authors are not aware of any affiliations, memberships, funding, or financial holdings that might be perceived as affecting the objectivity of this review. 

\section*{Acknowledgments} We thank Geoff Vallis for encouraging us to pursue this work. 
ADG is supported by the UK Engineering and Physical Sciences Research Council  (grant EP/T023139/1). JV is supported by the UK Natural Environment Research Council (grant NE/W002876/1).  For the purpose of open access, the authors have applied a Creative Commons Attribution (CC BY) licence to any Author Accepted Manuscript version arising.
Data access statement: no data was produced or analysed in this article. 

%

\bibliographystyle{plainnat}
\bibliography{JVbib}

\begin{thebibliography}{58}
\providecommand{\natexlab}[1]{#1}
\providecommand{\url}[1]{\texttt{#1}}
\expandafter\ifx\csname urlstyle\endcsname\relax
  \providecommand{\doi}[1]{doi: #1}\else
  \providecommand{\doi}{doi: \begingroup \urlstyle{rm}\Url}\fi

\bibitem[Andrews and McIntyre(1978{\natexlab{a}})]{andr-mcin78a}
D.~G. Andrews and M.~E. McIntyre.
\newblock An exact theory of nonlinear waves on a {L}agrangian-mean flow.
\newblock \emph{J. Fluid Mech.}, 89:\penalty0 609--646, 1978{\natexlab{a}}.

\bibitem[Andrews and McIntyre(1978{\natexlab{b}})]{andr-mcin78b}
D.~G. Andrews and M.~E. McIntyre.
\newblock On wave-action and its relatives.
\newblock \emph{J. Fluid Mech.}, 89:\penalty0 647--664, 1978{\natexlab{b}}.
\newblock .\ Addendum \textit{ibid.} \textbf{95}, 796.

\bibitem[Arnold(1966)]{arno66}
V.~I. Arnold.
\newblock Sur la g{\'e}om{\'e}trie diff\'erentielle des groupes de {L}ie de
  dimension infinie et ses applications {\`a} l'hydrodynamique des fluides
  parfaits.
\newblock \emph{Ann. Inst. Fourier}, 16:\penalty0 316--361, 1966.

\bibitem[Arnold and Khesin(1998)]{arno-kesh}
V.~I. Arnold and B.~A. Khesin.
\newblock \emph{Topological methods in hydrodynamics}, volume 125 of
  \emph{Applied mathematical sciences}.
\newblock Springer, 1998.

\bibitem[Bretherton and Schär(1993)]{bret-scha93}
C.~S. Bretherton and C.~Schär.
\newblock Flux of potential vorticity substance: A simple derivation and a
  uniqueness property.
\newblock \emph{J. Atmos. Sci.}, 50:\penalty0 1834--1836, 1993.

\bibitem[Bretherton(1971)]{bret71}
F.~P. Bretherton.
\newblock The general linearized theory of wave propagation.
\newblock In \emph{Mathematical Problems in the Geophysical Sciences},
  volume~13 of \emph{Lect. Appl. Math.}, pages 61--102. Am. Math. Soc., 1971.

\bibitem[B\"uhler(2014)]{buhl14}
O.~B\"uhler.
\newblock \emph{Waves and mean flows}.
\newblock Cambridge University Press, 2nd edition, 2014.

\bibitem[B\"uhler and McIntyre(1998)]{buhl-mcin98}
O.~B\"uhler and M.~E. McIntyre.
\newblock On non-dissipative wave--mean interactions in the atmosphere or
  oceans.
\newblock \emph{J. Fluid Mech.}, 354:\penalty0 301--343, 1998.

\bibitem[B\"uhler and McIntyre(2005)]{buhl-mcin05}
O.~B\"uhler and M.~E. McIntyre.
\newblock Wave capture and wave--vortex duality.
\newblock \emph{J. Fluid Mech.}, 534:\penalty0 67--95, 2005.

\bibitem[Craik and Leibovich(1976)]{crai-leib76}
A.~D. Craik and S.~Leibovich.
\newblock A rational model for {L}angmuir circulations.
\newblock \emph{J. Fluid Mech.}, 73:\penalty0 401--426, 1976.

\bibitem[Dewar(1970)]{dewa70}
R.~L. Dewar.
\newblock Interaction between hydromagnetic waves and a time-dependent
  inhomogeneous medium.
\newblock \emph{Phys. Fluids}, 13:\penalty0 2710--2720, 1970.

\bibitem[Ebin and Marsden(1970)]{ebin-mars}
D.~G. Ebin and J.~E. Marsden.
\newblock Groups of diffeomorphisms and the motion of an incompressible fluid.
\newblock \emph{Ann. Math.}, 92:\penalty0 102--163, 1970.

\bibitem[Eckart(1963)]{ecka63}
C.~Eckart.
\newblock Some transformations of the hydrodynamic equations.
\newblock \emph{Phys. Fluids}, 6:\penalty0 1037--1041, 1963.

\bibitem[Frankel(2004)]{fran04}
T.~Frankel.
\newblock \emph{The geometry of physics}.
\newblock Cambridge University Press, 2nd edition, 2004.

\bibitem[Gilbert and Vanneste(2018)]{gilb-v18}
A.~D. Gilbert and J.~Vanneste.
\newblock Geometric generalised {L}agrangian-mean theories.
\newblock \emph{J. Fluid Mech.}, 839:\penalty0 95--134, 2018.

\bibitem[Gilbert and Vanneste(2021)]{gilb-v21}
A.~D. Gilbert and J.~Vanneste.
\newblock A geometric look at {MHD} and the {B}raginski dynamo.
\newblock \emph{Geophys. Astrophys. Fluid Dynam.}, 115:\penalty0 436--471,
  2021.

\bibitem[Gjaja and Holm(1999)]{gjaj-holm}
I.~Gjaja and D.~D. Holm.
\newblock Self-consistent {H}amiltonian dynamics of wave mean-flow interaction
  for a rotating stratified incompressible fluid.
\newblock \emph{Physica D}, 98:\penalty0 343--378, 1999.

\bibitem[Grimshaw(1975)]{grim75}
R~Grimshaw.
\newblock Nonlinear internal gravity waves in a rotating fluid.
\newblock \emph{J. Fluid Mech.}, 71\penalty0 (3):\penalty0 497--512, 1975.

\bibitem[Grimshaw(1984)]{grim84}
R.~Grimshaw.
\newblock Wave action and wave--mean flow interaction, with application to
  stratified shear flows.
\newblock \emph{Ann. Rev. Fluid Mech.}, 16:\penalty0 11--44, 1984.

\bibitem[Haynes and McIntyre(1987)]{hayn-mcin87}
P.~H. Haynes and M.~E. McIntyre.
\newblock On the evolution of vorticity and potential vorticity in the presence
  of diabatic heating and frictional or other forces.
\newblock \emph{J. Atmos. Sci.}, 44:\penalty0 828--841, 1987.

\bibitem[Haynes and McIntyre(1990)]{hayn-mcin90}
P.~H. Haynes and M.~E. McIntyre.
\newblock On the conservation and impermeability theorems for potential
  vorticity.
\newblock \emph{J. Atmos. Sci.}, 47:\penalty0 2021--2031, 1990.

\bibitem[Holm(1996)]{holm96}
D.~D. Holm.
\newblock The ideal {C}raik--{L}eibovich equations.
\newblock \emph{Physica D}, 98:\penalty0 415--449, 1996.

\bibitem[Holm(1999)]{holm99}
D.~D. Holm.
\newblock Fluctuation effects on 3{D} {L}agrangian mean and {E}ulerian mean
  fluid motion.
\newblock \emph{Physica D}, 133:\penalty0 215--269, 1999.

\bibitem[Holm(2002{\natexlab{a}})]{holm02a}
D.~D. Holm.
\newblock Variational principles for {L}agrangian-averaged fluid dynamics.
\newblock \emph{J. Phys. A. Math. Gen.}, 35:\penalty0 679--668,
  2002{\natexlab{a}}.

\bibitem[Holm(2002{\natexlab{b}})]{holm02b}
D.~D. Holm.
\newblock Lagrangian averages, averaged {L}agrangians, and the mean effects of
  fluctuation in fluid dynamics.
\newblock \emph{Chaos}, 12:\penalty0 518--530, 2002{\natexlab{b}}.

\bibitem[Holm(2019)]{holm19}
D.~D. Holm.
\newblock Stochastic closures for wave--current interaction dynamics.
\newblock \emph{J. Nonlinear Sci.}, 29:\penalty0 2987--3031, 2019.

\bibitem[Holm(2021)]{holm21}
D.~D. Holm.
\newblock Stochastic variational formulations of fluid wave--current
  interaction.
\newblock \emph{J. Nonlinear Sci.}, 31:\penalty0 4, 2021.

\bibitem[Holm et~al.(1985)Holm, Marsden, Ratiu, and Weinstein]{holm-et-al}
D.~D. Holm, J.~E. Marsden, T.~Ratiu, and A.~Weinstein.
\newblock Nonlinear stability of fluid and plasma equilibria.
\newblock \emph{Phys. Rep.}, 123:\penalty0 1--116, 1985.

\bibitem[Holm et~al.(1998)Holm, Marsden, and Ratiu]{holm-et-al98}
D.~D. Holm, J.~E. Marsden, and T.~Ratiu.
\newblock The {E}uler--{P}oincar\'e equations and semi-direct products with
  applications to continuum theories.
\newblock \emph{Adv. Math.}, 137:\penalty0 1--81, 1998.

\bibitem[Holm et~al.(2023)Holm, Hu, and Street]{holm-hu-stree}
D.~D. Holm, R.~Hu, and O.~Street.
\newblock Lagrangian reduction and wave mean flow interaction.
\newblock \emph{Physica D}, 454:\penalty0 133847, 2023.

\bibitem[Holmes-Cerfon et~al.(2011)Holmes-Cerfon, {B{\"u}hler}, and
  Ferrari]{holm-et-al11}
M.~Holmes-Cerfon, O.~{B{\"u}hler}, and R.~Ferrari.
\newblock Particle dispersion by random waves in the rotating {B}oussinesq
  system.
\newblock \emph{J. Fluid Mech.}, 670:\penalty0 150--175, 2011.

\bibitem[Kafiabad(2022)]{kafi22}
H.~A. Kafiabad.
\newblock Grid-based calculation of the {L}agrangian mean.
\newblock \emph{J. Fluid Mech.}, 940:\penalty0 A21, 2022.

\bibitem[Kafiabad and Vanneste(2023)]{kafi-v23}
H.~A. Kafiabad and J.~Vanneste.
\newblock Computing {L}agrangian means.
\newblock \emph{J. Fluid Mech.}, 960:\penalty0 A36, 2023.

\bibitem[Kafiabad et~al.(2021)Kafiabad, Vanneste, and Young]{kafi-et-al21}
H.~A. Kafiabad, J.~Vanneste, and W.~R. Young.
\newblock Wave-averaged balance: a simple example.
\newblock \emph{J. Fluid Mech.}, 911:\penalty0 R1, 2021.

\bibitem[Leibovich(1976)]{leib80}
S.~Leibovich.
\newblock On wave-current interaction theories of {L}angmuir circulations.
\newblock \emph{J. Fluid Mech.}, 99:\penalty0 715--724, 1976.

\bibitem[Lichtenberg and Lieberman(1992)]{lich-lieb}
A.J. Lichtenberg and M.A. Lieberman.
\newblock \emph{Regular and stochastic motion}.
\newblock Springer--Verlag, 2nd edition, 1992.

\bibitem[Marsden and Shkoller(2001)]{mars-shko01}
J.E. Marsden and S.~Shkoller.
\newblock Global well-posedness for the {L}agrangian averaged
  {N}avier--{S}tokes ({LANS}-$\alpha$) equations on bounded domains.
\newblock \emph{Phil. Trans. R. Soc. Lond. A}, 359:\penalty0 1449--1468, 2001.

\bibitem[Marsden and Shkoller(2003)]{mars-shko03}
J.E. Marsden and S.~Shkoller.
\newblock The anisotropic {L}agrangian averaged {E}uler and {N}avier--{S}tokes
  equations.
\newblock \emph{Arch. Ration. Mech. Anal.}, 166:\penalty0 27--46, 2003.

\bibitem[McIntyre(1988)]{mcin88}
M.~E. McIntyre.
\newblock A note on the divergence effect and the {L}agrangian-mean surface
  elevation in periodic water waves.
\newblock \emph{J. Fluid Mech.}, 189:\penalty0 235--242, 1988.

\bibitem[Moore(1970)]{moor70}
D.~Moore.
\newblock The mass transport velocity induced by free oscillations at a single
  frequency.
\newblock \emph{Geophysical Fluid Dynamics}, 1:\penalty0 237--247, 1970.

\bibitem[Oliver and Vasylkevych(2019)]{oliv-vasy}
M.~Oliver and S.~Vasylkevych.
\newblock Geodesic motion on groups of diffeomorphisms with {H1} metric as
  geometric generalised {L}agrangian mean theory.
\newblock \emph{Geophys. Astrophys. Fluid Dynam.}, 113:\penalty0 466--490,
  2019.

\bibitem[Roberts and Soward(2006{\natexlab{a}})]{robe-sowa06a}
P.~H. Roberts and A.~M. Soward.
\newblock Eulerian--{L}agrangian means in rotating magnetohydrodynamic flows
  {I}. general results.
\newblock \emph{Geophys. Astrophys. Fluid Dynam.}, 100:\penalty0 457--483,
  2006{\natexlab{a}}.

\bibitem[Roberts and Soward(2006{\natexlab{b}})]{robe-sowa06b}
P.~H. Roberts and A.~M. Soward.
\newblock Covariant description of non-relativistic magnetohydrodynamics.
\newblock \emph{Geophys. Astrophys. Fluid Dynam.}, 100:\penalty0 457--483,
  2006{\natexlab{b}}.

\bibitem[Roberts and Soward(2009)]{robe-sowa09}
P.~H. Roberts and A.~M. Soward.
\newblock The {N}avier--{S}tokes-$\alpha$ equations revisited.
\newblock \emph{Geophys. Astrophys. Fluid Dynam.}, 103:\penalty0 303--316,
  2009.

\bibitem[Salmon(1998)]{salm98}
R.~Salmon.
\newblock \emph{Lectures on geophysical fluid dynamics}.
\newblock Oxford University Press, 1998.

\bibitem[Salmon(2013)]{salm13}
Rick Salmon.
\newblock An alternative view of generalized {L}agrangian mean theory.
\newblock \emph{J. Fluid Mech.}, 719:\penalty0 165--182, 2013.

\bibitem[Salmon(2016)]{salm16}
Rick Salmon.
\newblock Variational treatment of inertia--gravity waves interacting with a
  quasi-geostrophic mean flow.
\newblock \emph{J. Fluid Mech.}, 809:\penalty0 502--529, 2016.

\bibitem[Schutz(1980)]{schu80}
B.~Schutz.
\newblock \emph{Geometrical methods of mathematical physics}.
\newblock Cambridge University Press, 1980.

\bibitem[Soward(1972)]{sowa72}
A.~M. Soward.
\newblock A kinematic theory of large magnetic {R}eynolds number dynamos.
\newblock \emph{Phil. Trans. R. Soc. London}, A272:\penalty0 431--462, 1972.

\bibitem[Soward and Roberts(2008)]{sowa-robe08}
A.~M. Soward and P.~H. Roberts.
\newblock On the derivation of the {N}avier--{S}tokes--$\alpha$ equations from
  {H}amilton’s principle.
\newblock \emph{J. Fluid Mech.}, 604:\penalty0 297--323, 2008.

\bibitem[Soward and Roberts(2010)]{sowa-robe10}
A.~M. Soward and P.~H. Roberts.
\newblock The hybrid {E}uler--{L}agrange procedure using an extension of
  {M}offatt’s method.
\newblock \emph{J. Fluid Mech.}, 661:\penalty0 45--72, 2010.

\bibitem[Soward and Roberts(2014)]{sowa-robe14}
A.~M. Soward and P.~H. Roberts.
\newblock Eulerian--{L}agrangian means in rotating magnetohydrodynamic flows
  {II}. {B}raginsky’s nearly axisymmetric dynamo.
\newblock \emph{Geophys. Astrophys. Fluid Dynam.}, 108:\penalty0 269--322,
  2014.

\bibitem[Suzuki and Fox-Kemper(2016)]{suzu-foxk}
N.~Suzuki and B.~Fox-Kemper.
\newblock Understanding {S}tokes forces in the wave-averaged equations.
\newblock \emph{J. Geophys. Res.}, C121:\penalty0 3579--3596, 2016.

\bibitem[Tao(2016)]{tao16}
T.~Tao.
\newblock Finite time blowup for {L}agrangian modifications of the
  three-dimensional {E}uler equation.
\newblock \emph{Ann. PDE}, 2:\penalty0 9, 2016.

\bibitem[Vallis(2006)]{vall17}
G.~K. Vallis.
\newblock \emph{Atmospheric and oceanic fluid dynamics: fundamentals and
  large-scale circulation}.
\newblock Cambridge University Press, 2nd edition, 2006.

\bibitem[Vanneste and Young(2022)]{v-youn22}
J.~Vanneste and W.~R. Young.
\newblock Stokes drift and its discontents.
\newblock \emph{Phil. Trans. R. Soc. Lond. A}, 380:\penalty0 20210032, 2022.

\bibitem[Wagner and Young(2015)]{wagn-youn}
G.~L. Wagner and W.~R. Young.
\newblock Available potential vorticity and wave-averaged quasi-geostrophic
  flow.
\newblock \emph{J. Fluid Mech.}, 785:\penalty0 401--424, 2015.

\bibitem[Xie and Vanneste(2015)]{xie-v15}
J.-H. Xie and J.~Vanneste.
\newblock A generalised-{L}agrangian-mean model of the interactions between
  near-inertial waves and mean flow.
\newblock \emph{J. Fluid Mech.}, 774:\penalty0 143--169, 2015.

\end{thebibliography}

\end{document}